\documentclass[twocolumn,pre,floatfix,superscriptaddress,longbibliography]{revtex4}
\usepackage{amsbsy}
\usepackage{latexsym,epsfig,graphicx}
\usepackage{dcolumn}
\usepackage{graphicx}
\usepackage{subfigure}
\usepackage{comment}
\usepackage{color}
\usepackage{bm}
\usepackage{mathrsfs}
\usepackage{amsfonts}
\usepackage{amsmath}
\usepackage{color}
\usepackage{amssymb}
\usepackage{xspace}
\usepackage{epstopdf}
\usepackage{tabularx}
\usepackage{longtable}
\usepackage[colorlinks=true, letterpaper=true, pdfstartview=FitV, linkcolor=blue, citecolor=blue, urlcolor=blue]{hyperref}
\usepackage[normalem]{ulem}
\usepackage{multirow}
\usepackage{esint}
\usepackage{mathdots}

\allowdisplaybreaks[4]
\pdfoutput=1

\begin{document}

\title{Direct comparison of stochastic-driven nonlinear dynamical systems for combinatorial optimization}

\author{Junpeng Hou}
\thanks{The work was performed when J.~H.~was an intern at Microsoft.}
\affiliation{Pinterest Inc., San Francisco, California 94103, USA}

\author{Amin Barzegar}
\affiliation{Microsoft Quantum, Microsoft, Redmond, Washington 98052, USA}

\author{Helmut G. Katzgraber}
\thanks{The work of H.~G.~K.~was performed before joining Amazon.}
\affiliation{Amazon Advanced Solutions Lab, Seattle, Washington 98170, USA}

\begin{abstract}

Combinatorial optimization problems are ubiquitous in industrial applications. However, finding optimal or close-to-optimal solutions can often be extremely hard. Because some of these problems can be mapped to the ground-state search of the Ising model, tremendous effort has been devoted to developing solvers for Ising-type problems over the past decades. Recent advances in controlling and manipulating both quantum and classical systems have enabled novel computing paradigms such as quantum simulators and coherent Ising machines to tackle hard optimization problems. Here, we examine and benchmark several physics-inspired optimization algorithms, including coherent Ising machines, gain-dissipative algorithms, simulated bifurcation machines, and Hopfield neural networks, which we collectively refer to as stochastic-driven nonlinear dynamical systems. Most importantly, we benchmark these algorithms against random Ising problems with planted solutions and compare them to simulated annealing as a baseline leveraging the same software stack for all solvers. We further study how different numerical integration techniques and graph connectivity affect performance. This work provides an overview of a diverse set of new optimization paradigms.

\end{abstract}

\maketitle

\section{Introduction}

The ever-growing need for computing has surged over the past few decades as we are facing unprecedentedly complex problems in modern society, industry, and science. Among them, combinatorial optimization problems stand out for being notoriously difficult to solve as a result of the so-called \textit{combinatorial explosion} \cite{arora:09}, meaning that their solution space grows exponentially with increasing number of variables. Consequently, many combinatorial optimization problems are known to be nondeterministic polynomial (NP)-hard or NP-complete. Naturally, these problems have attracted considerable attention from various fields such as computational science \cite{smith:99} and operations research \cite{juan:15}.

From a physical perspective, it is known that many combinatorial optimization problems can be mapped to the problem of finding the ground states of the Ising model, and such mapping can be achieved using polynomial resources \cite{barahona:82}. Note, however, that the overhead can be sizable in some cases. Several examples, such as the mapping for the traveling salesperson problem, the knapsack problem, or maximum independent set problem can be found in Ref.~\cite{lucas:14}. Moreover, the binary variables $\sigma_l=\pm1$ can be modeled through a plethora of physical systems including classical and quantum spins \cite{kirkpatrick:83,boettcher:01,wang:15c,sutton:17,kadowaki:98,farhi:00,leleu:19}, optical fields \cite{wang:13b,marandi:14,inagaki:16a,clements:17,pierangeli:19}, solid-state circuits \cite{mahboob:16,sheldon:19,wang:19a,chou:19,cai:20,afoakwa:20}, and nonequilibrium condensates \cite{berloff:17, kalinin:18a}. These algorithms are heuristics that start from a random initial state. The probability to find the ground state (or the lowest-energy state) depends highly on whether or not the system's dynamics can escape from local minima effectively. The energy landscape of many combinatorial optimization problems is generally rather rough with exponentially many metastable states \cite{leleu:19}. Therefore, the resources required by heuristics to reach the low-lying states scale nonpolynomially with the problem size.

More recently, various new computing paradigms have been proposed and realized in different physical systems to more effectively traverse energy landscapes by allowing the original binary variable $\sigma_l=\pm1$ to relax to a continuous variable of either real $x_l\in\mathbb{R}$ or complex $\Psi_l\in\mathbb{C}$ type. In these algorithms, optimization is achieved by mapping local minima to fixed points through bifurcation processes which are usually created by tuning a set of gain/dissipative strengths (or equivalently, proper driven effects) in the system \cite{Strogatz:15}. Throughout this work, we refer to these physical systems as \textit{stochastic-driven nonlinear dynamical systems} (SDNDSs), because the name highlights their three common key features, namely, stochastic noise terms, driven or gain/dissipative effects, and nonlinearity.

In fact, similar ideas can be traced back to early work on Hopfield neural networks in which the continuous variable of each neuron is mapped to a binary variable via an activation function \cite{hopfield:82}. It has also been shown that reducing the gain or the steepness of the activation function could lead to an exponential decrease in the number of local minima, and therefore, improve the solution quality \cite{hopfield:85,waugh:90,fukai:90}. Despite early skepticism \cite{wilson:88}, more recent studies have argued that such algorithms might outperform state-of-the-art heuristics or perform equally well at the very least \cite{leleu:17,leleu:19}.

To address this outstanding debate and to gain a more comprehensive understanding of SDNDSs, in this work we examine some representative systems and algorithms using the same code base, including coherent Ising machines (CIMs), gain-dissipative (GD) platforms, oscillator-based Ising machines (OIM), simulated bifurcation machines (SBM), as well as classical Hopfield neural networks. We benchmark these algorithms using Ising problems with planted solutions and compare their performances to that of simulated annealing (SA) \cite{kirkpatrick:83}, which is currently a popular and efficient heuristic for binary combinatorial optimization problems \cite{isakov:15}.

Furthermore, we conduct a range of experiments to deeper investigate the performance of these heuristics. First, because the dynamical systems studied here are simulated by numerically solving the associated equations of motion (EoM) which are usually ordinary differential equations (ODEs),  we study how using different numerical integration methods---e.g., Runge–Kutta (RK)---affect  performance. To better understand how graph connectivity affects performance, we systematically dilute a fully-connected Sherrington-Kirkpatrick model \cite{sherrington:75} to a Viana-Bray model \cite{viana:85}. To this end, we compare the performance of SA against the SBM.

\begin{table}[t!]
\centering
\caption{List of different stochastic-driven nonlinear dynamical systems studied in this work. The artificial spin is denoted by either $x_l$ (real field) or $\Psi_l$ (complex field). Other common parameters include the pumping/driving strength $\alpha$, global Ising coupling $\beta$, and the nonlinear activation function $\phi(\bullet)$. For other model-specific parameters, please refer to the relevant subsections in Sec.~\ref{Models}.}
 \vspace{1.5mm}
\begin{tabular*}{\columnwidth}{c l} 
 \hline
 \hline
 Algorithm &  Equations of motion\\
 \hline
 \vspace{1mm}\\
OPO-CIM & $\dot{x}_l=\alpha x_l-\frac{2}{3}x_l^3+\beta\sum\limits_mJ_{lm}x_m+g_l$
\vspace{1mm}\\
 Sim-CIM & $x_l\leftarrow \phi\left((\alpha+1) x_l+\beta\sum\limits_mJ_{lm}x_m\right)+g_l$
\vspace{3mm}\\
 OEO-CIM & $x_l\leftarrow \cos^2(\alpha x_l+\beta\sum\limits_mJ_{lm}x_m-\frac{\pi}{4})+g_l-\frac{1}{2}$
\vspace{2mm}\\
 OIM & $\dot{x}_l=-\alpha\sin(2x_l)-\beta\sum\limits_mJ_{lm}\sin(x_l-x_m)+g_l$
\vspace{4mm}\\
 Hopfield & $\dot{x}_l=-\alpha x_l+\beta\sum\limits_mJ_{lm}\phi\left(x_m\right)+g_l$
\vspace{2mm}\\
 GD & $\begin{cases} 
 \dot{\Psi}_l=\Psi_l\left(\alpha_l-\alpha_{c0}-|\Psi_l|^2\right)\\
    \hspace{0.8cm}+\sum\limits_{m}\beta_{lm}K_{lm}\Psi_m+\sum\limits_{q=1}^2h_{ql}\Psi_l^*+g_l,\\
 \dot{\alpha}_l=\epsilon\left(\rho_{\text{th}}-|\Psi_l|^2\right), \vspace{1mm}\\ \dot{K}_{lm}=\epsilon'\left(J_{lm}-\beta_{lm}K_{lm}\right)
 \end{cases}$
\vspace{3mm}\\
  SBM & $\begin{cases} 
 \dot{x}_l=\Delta y_l,  \vspace{2mm}\\
 \dot{y}_l=-\left(Kx_l^2-\alpha(t)+\Delta \right)x_l+\beta\sum\limits_mJ_{lm}x_m
 \end{cases}$
 \vspace{1mm}\\
 \hline
 \hline
\end{tabular*}
\label{Table1: Summary of models}
\end{table}

\section{Overview of physical models}
\label{Models}

In this section, we give a brief overview of the SDNDSs studied. We present the underlying physics of these algorithms and make comparisons between them. Besides the recently proposed SDNDSs, such as different Ising machines and gain-dissipative methods, we also discuss the Hopfield neural network, which is intrinsically a nonlinear dynamical system. For the convenience of the reader, we have summarized all the SDNDSs studied here in Table~\ref{Table1: Summary of models}. Throughout this work, we adopt a consistent notation for the  corresponding EoMs.

All the SDNDSs are designed to optimize Ising-like Hamiltonians (which are related to quadratic-unconstrained binary optimization problems) given by the following Hamiltonian:
\begin{equation}
    H = -\frac{1}{2}\sum_{l,m}J_{lm}\sigma_l\sigma_m - \sum_lh_l\sigma_l \,.
\end{equation}
Here, the spin-spin interaction matrix $J$ is self-adjoint, namely, $J^*_{lm} = J_{ml}$ (because Hamiltonian dynamics are hermitian) and $h_l$ is a real onsite magnetic field. For practical purposes, we assume that the interaction matrix is real and therefore symmetric. Moreover, without loss of generality, we consider only the Ising model with no magnetic field, i.e. $h_l=0$, because an $N$-spin Ising model with a magnetic field can be mapped to an $(N+1)$-spin Ising model without one using polynomial resources \cite{goto:21}.

\subsection{Coherent Ising machine and its variations} 
\label{CIM}

Physical Ising machines usually suffers from a limited number of artificial spins when operating in the quantum regime. However, a classical system like an optical platform has been used to demonstrate the so-called coherent Ising machine (CIM), which can tackle optimization problems with thousands of nodes \cite{inagaki:16b}. The CIM architecture is primarily based on an optical parametric oscillator (OPO) consisting of a fiber loop and a degenerate parametric amplifier or a single-mode squeezer. The OPO is pumped above its oscillation threshold, hence exciting the pulse modes circulating inside. Thanks to the phase sensitivity of the optical squeezer, the phases of these modes tend to become locked at either $0$ or $\pi$, giving rise to a binary constraint.

CIMs can be regarded as optical neural networks operating in the quantum limit, yet at room temperature. Consequently, CIMs may outperform some classical heuristics like SA \cite{yamamoto:17} for some quadratic binary optimization problems. CIMs have been benchmarked against the D-Wave quantum annealer \cite{hamerly:19}, as well as classical neural networks \cite{haribara:17}. In both cases, the CIM showed better performance although the merits are highly problem-dependent.

Given the success of the bare-bones OPO-CIM, many different optical Ising machines have been proposed. For example, a recent work utilizes spatial light modulation such that the setup can be scaled up more easily \cite{pierangeli:19}. In this paper, we focus on the following two variants of the OPO-CIM model.

\subsubsection{Simulated coherent Ising machine}

The simulated CIM or Sim-CIM was proposed as a variant of the CIM that suits fast simulations \cite{tiunov:19}. In the Sim-CIM, each pulse inside the OPO is written as $(x_l+ip_l)/\sqrt{2}$ in  canonical coordinates. By deriving the per-round change of each pulse and ignoring nonlinear losses, the real part of the EoM reads
\begin{equation}
    \Delta x_l = \alpha x_l + \beta \sum_mJ_{lm}x_m+g_l,
    \label{Sim-CIM EoM}
\end{equation}
where $\alpha$ denotes the linear pumping or gain strength, $\beta$ is the global Ising coupling strength and $g_l$ represents a site-dependent noise term. We note that the three terms on the right-hand side of Eq.~\eqref{Sim-CIM EoM} play important roles in a range of dynamical systems studied in this work. Therefore, the same notation is used for all SDNDSs throughout the paper. Note also that most of the parameters are time-dependent (except the Ising coupling $J_{lm}$). However, we omit this time dependence in our notation for better readability.

Guided by the Hopfield method (see also Sec.~\ref{HNN}), the ``spins'' $x_l$ are updated via an activation function $\phi(\bullet)$ to account for saturation
\begin{equation}
    x_l\leftarrow\phi(x_l+\Delta x_l).
\end{equation}
A recent study has shown that the choice of a nonlinear activation function can significantly affect the performance of Ising machines by influencing the homogeneity of their spins \cite{boehm:19}. Here, we follow the original proposal and use a hyperbolic tangent profile for the pumping term $\alpha$ and a discontinuous activation function of the form
\begin{equation}
    \phi(x)=
        \begin{cases}
            x & \quad \text{if } |x|\leq x_\text{sat} \\
            x_\text{sat}  & \quad \text{otherwise} , 
        \end{cases},
\end{equation}
where $x_\text{sat}$ defines a hard-wall boundary and confines the spins within the interval $[-x_\text{sat}, x_\text{sat}]$.

\subsubsection{Optoelectronic oscillator coherent Ising machine}

Due to the phase sensitivity of the generation, interference, and detection of coherent optical states, stable operations can be challenging in the OPO-CIM. To overcome this shortcoming, a fully programmable CIM or the optoelectronic oscillator (OEO) CIM was proposed \cite{boehm:19}.

The OEO design consists of both an optical and an electrical pathway. The optical part feeds the output of a laser diode to a Mach–Zehnder modulator, which is then passed to a photo diode. The electrical pathway creates feedback by sampling the photo voltage. Such an OEO feedback system is nonlinear in nature and the dynamics are described by:
\begin{equation}
    x_l \leftarrow \cos^2\left(f_l-\frac{\pi}{4}+g_l\right)-\frac{1}{2},
\end{equation}
where the feedback term $f_l$ is similar to that of the CIM or Sim-CIM
\begin{equation}
    f_l = \alpha x_l+\beta \sum_mJ_{lm}x_m.
\end{equation}
In this dynamical system, a rather unique activation function, i.e., $\cos^2(\bullet)$ appears. This is justified by the Taylor expansion near $x_l=0$ because the spin amplitude is small when the system is driven to the laser threshold ($\alpha\approx1$) where CIMs have their optimal performance \cite{leleu:17}. Assuming that the pumping term is dominant over the Ising potential, namely, $\alpha x_l\gg\beta \sum_mJ_{lm}x_m$ and truncating the corresponding Taylor series at the third order, we can easily recover the equations of motion of the OPO-CIM from those of the OEO-CIM (see Table~\ref{Table1: Summary of models}). The Sim-CIM is also similar to OEO-CIM except with a different activation function.

\subsection{Gain-dissipative platforms}
\label{GD}

The Gain-dissipative algorithms are inspired by the nonequilibrium dynamics of a polariton Bose-Einstein condensate \cite{kalinin:18a,kalinin:18b}. While, in principle, the discussed CIM models (or the oscillator-based Ising machine to be discussed in Sec.~\ref{OIM}) can also be regarded as a gain-dissipative system because of the pumping gain and/or material loss in the optical systems, in this work, we use the term ``GD'' to refer specifically to the algorithm stemming from the non-equilibrium dynamics of an exciton-polariton condensate. 

The Gross–Pitaevskii equation \cite{landau:80}, which governs the collective dynamics of the aforementioned polariton Bose-Einstein condensate, reads
\begin{align}
    \dot{\Psi}_l=&\,\Psi_l\left(\alpha_l-\alpha_{c0}-|\Psi_l|^2\right)+\sum_{m}\beta_{lm}K_{lm}\Psi_m\nonumber\\
    &+\sum_{q=1}^2h_{ql}\Psi_l^*+g_l,\label{GD EoM}
\end{align}
where $\Psi_l$, $\alpha_l$ and $\alpha_{c0}$ represent the complex order parameter of the condensate and the particle injection and loss rates, respectively. Note that while $\alpha_l$ is site dependent, $\alpha_{c0}$ is uniform across all sites. For comparison, a uniform $\alpha_l$ and $\alpha_{c0}$ are similar to the parametric gain $w_\alpha$ and linear loss $\gamma_\alpha$ in the Sim-CIM model, reparameterized as $\alpha = w_\alpha-\gamma_\alpha$.

The coefficients $\beta_{lm}K_{lm}$ represent the coupling strength between $\Psi_l$ and $\Psi_m$ in which $\beta_{lm}$ arises from particle injection effects, whereas $K_{lm}$ is induced by other mechanisms like bare hopping or the laser-controlled coupling \cite{berloff:17}. Unlike the global Ising coupling $\beta$ in other models, GD systems use a tunable local coupling strength $\beta_{lm}$ instead (see Table~\ref{Table1: Summary of models}). Physically, a common choice for the coupling coefficients is $\beta_{lm}=\alpha_l+\alpha_m$, where the onsite injection rate $\alpha_l$ is governed by the following dynamics:
\begin{equation}
    \dot{\alpha}_l=\epsilon\left(\rho_{\text{th}}-|\Psi_l|^2\right).
\end{equation}
This ensures that the network can be synchronized to reach a uniform density distribution of $\rho_{\text{th}}$ after some equilibration time. The constant $\epsilon$ controls how fast such synchronization will occur. Because the local Ising coupling strength $\beta_{lm}$ can depend on the injection rates, the coupling $K_{lm}$ is controlled via
\begin{equation}
    \dot{K}_{lm}=\epsilon'\left(J_{lm} - \beta_{lm}K_{lm}\right),
\end{equation}
such that the Ising coupling $J_{lm}$ is recovered at the fixed points.

Unlike the two CIM models discussed earlier, the GD platform is highly versatile, in the sense that it can be tuned to suit different types of optimization problems. For example, if one sets $h_{ql}=0$ in Eq.~\eqref{GD EoM}, the GD model can be used to optimize the XY Hamiltonian \cite{yeomans:92}. However, if $h_{2m} > \sum_{l\neq m}|J_{lm}|$, the phases will be fixed at $0$ or $\pi$, and therefore, a binary variable problem can be simulated. With other values of $h_{ql}$, the algorithm can fit a $q$-state clock model in which the phases are forced to be $2\pi n/q,n\in\mathbb{Z}$ \cite{kalinin:18b}. The GD system's adaptability can also be exploited in other nontrivial ways. For instance, a recent work demonstrated a physics-inspired algorithm that solves higher-order polynomial binary optimization problems without mapping them to a quadratic unconstrained binary form \cite{Kalinin:21}. However, higher-order terms become more difficult to numerically integrate and might lead to numerical instabilities. As another example, by breaking the $U(1)$ symmetry of the order parameter into $\mathbb{Z}_3$, a ternary spin system can be introduced to tackle the Max-3-SAT problem \cite{harrison:20} natively.

\subsection{Oscillator-based Ising machines} 
\label{OIM}

More recently, a new way to engineer Ising machines using a network of coupled self-sustaining nonlinear oscillators called OIM was proposed \cite{wang:19a}. Unlike CIMs, the system can be experimentally realized by conventional CMOS electronics, making it more cost- and energy-efficient.

The OIM utilizes subharmonic injection locking (SHIL) to lock the oscillators at discrete phases. Specifically, when a second-harmonic signal (a perturbation with double the natural frequency of a single oscillator) is injected into every oscillator in the network, all the oscillator phases will become binary under SHIL. The EoM of this dynamical system is
\begin{equation}
    \dot{x}_l=-\alpha\sin(2x_l)-\beta\sum_mJ_{lm}\sin(x_l-x_m)+g_l,
    \label{EoM-OIM}
\end{equation}
where $\alpha$ is the coupling strength from the second harmonic signal, and the parameter $\beta$ modulates the overall coupling strength of the network, similar to other systems discussed so far. 

Similar to CIMs, a possible fixed point of OIMs is not guaranteed to be the global optimum. This is because the global Lyapunov function of such a coupled oscillator network has local minima and, in general, the dynamical system can converge to a fixed point that corresponds to a suboptimal solution. While a careful selection of the noise strength $|g_l|$ can prevent the OIM from getting trapped in local minima, optimized annealing schedules for all three parameters $\alpha$, $\beta$, and $g_l$ are crucial to its overall performance \cite{wang:19a}.

The OIM can be easily realized using LC oscillators as first demonstrated by Wang \emph{et al.}~\cite{wang:19b} as a small network of eight ``spins'' on a breadboard. An OIM consisting of up to $240$ oscillators has also been built on printed circuit boards (PCBs) that supports programmable couplings \cite{wang:19b}. Chou \emph{et al.}~\cite{chou:19} demonstrated a smaller prototype circuit of four nodes but with a cross-bar architecture, therefore offering additional advantages such as access to higher bit precision and better scalability to larger systems and denser graphs. 

\subsection{Simulated bifurcation machine} 
\label{SBM}

In addition to the oscillator networks we have presented so far, a different approach of applying adiabatic optimization on Kerr-nonlinear parametric oscillators (KPOs) has recently been proposed \cite{goto:16,nigg:17}. Such system shares some similarities with CIMs, however it is based on quantum mechanics. Without providing too much mathematical details, we briefly explain how to tune the gain $\alpha$ and the global Ising coupling $\beta$ in the original KPO model. The parameter $\beta$ should be set to a small value so that the vacuum ground states of the underlying physical system are preserved. In turn, $\alpha$ is adiabatically increased from zero to some large value---in comparison to the characteristic energy scale of the system---so that a cat state (an equal superposition of two coherent states with opposite parities) can be achieved in a single decoupled KPO. Eventually, each KPO is pumped to a coherent state through quantum adiabatic bifurcation, yielding the ground-state solution of the Ising problem. One advantage this system has over others is that both quadratic and quartic spin-spin interactions can be implemented. 

The SBM is obtained from a mean-field treatment of the KPO system by writing the KPO annihilation operator as a complex expression $x_l+iy_l$. Thus, the EoM can be derived through the classical Hamiltonian formalism by separating real and imaginary parts, and then can be further simplified as \cite{goto:19}
\begin{eqnarray}
    \dot{x}_l&=&\Delta y_l, \\ \nonumber
    \dot{y}_l&=&-\left(Kx_l^2-\alpha+\Delta\right)x_l+\beta\sum_mJ_{lm}x_m.\label{EoM-SBM}
\end{eqnarray}
Here, $K$ denotes the inter-oscillator interaction strength, which eventually becomes an onsite but site-dependent potential. A similar phenomenon is realized in the GD model via the inter-atomic interaction $-|\Psi_l|^2$ and through the nonlinear loss function $-x_l^3$ in the OPO-CIM. In the Sim-CIM or OEO-CIM, the effect of such a term is implicitly replaced by a nonlinear activation function. Physically, a strong density interaction prevents any extra population from accumulating in a single state. Similarly, the activation function bounds the variables to be in $[-1,1]$.

The $\Delta$-term in Eq.~\eqref{EoM-SBM} represents detuning, namely, the difference between the frequency of each KPO and half of the pumping frequency, which is assumed to be uniform for all KPOs in the SBM, and thus, it appears along with the pumping and the self-energy terms. The detuning also affects how strongly $x_l$ is coupled to $y_l$. Unlike other systems, the EoMs are intrinsically second-order differential equations because they are derived via a Hamiltonian formalism.

More recently, two variants of SBM have been proposed \cite{goto:21}. In the first variant, the density interaction term is replaced by an activation function similar to Sim-CIM. This is a result of our previous observation that the effect of a physical density interaction is closely related to that of a bounded activation function. The other variant of the SBM applies an additional activation function, namely, $\text{sgn}(\bullet)$ when computing the Ising potential term to reduce analog errors. These, in turn, can be used either for speed or precision.

\subsection{Hopfield neural network} 
\label{HNN}

Neural networks represent a broad class of nonlinear dynamical systems \cite{du:14}. The Hopfield model, being the best-known dynamical system, operates in an unsupervised manner. The original proposal for the discrete Hopfield model reads \cite{hopfield:82}
\begin{equation}
    x_i\leftarrow\text{sgn}\left(\sum_lJ_{lm}x_m\right),
\end{equation}
which can be regarded as a derandomized simulated annealing algorithm \cite{haribara:17}. Note that the Sim-CIM can be reduced to such a system by merely dropping the pumping term and applying a steeper activation function. Later, a similar recurrent neural network was proposed, where the update rule of the variables reads \cite{hopfield:84}
\begin{equation}
    x_l\leftarrow \sum_{m\neq l}J_{lm}\phi(x_m)+\theta_l .
\end{equation}
Here, $\phi$ is a sigmoidal activation function and $\theta_l$ is a bias to the neuron. Such system has been recently realized through memristor circuits demonstrating a power-efficient combinatorial optimizer \cite{cai:20}. Correspondingly, a continuous-time Hopfield model can be defined as
\begin{equation}
    \dot{x}_l=\sum_{m}J_{lm}\phi(x_m)+\theta_l.
\end{equation}
In combinatorial optimization problems, the continuous model usually outperforms the discrete model due to a smoother energy landscape \cite{du:14}. The continuous model has been improved for optimization problems by incorporating nonlinear graded-response model neurons \cite{hopfield:86}, i.e., 
\begin{equation} \label{HNN_def}
    \dot{x}_l=-\alpha x_l+\beta\sum_mJ_{ml}\phi(x_m),
\end{equation}
where the pumping strength $\alpha$ and the global Ising coupling $\beta$ are tunable constants. The Hopfield model as presented in Eq.~\eqref{HNN_def} is closely related to the Sim-CIM except that the activation function is only applied to its neurons (spins) rather than the entire right-hand side of the EoM as done in the Sim-CIM model. Throughout this work, we refer to Eq.~\eqref{HNN_def} when we consider the Hopfield method unless specified otherwise.

\section{Benchmark results on Ising problems with planted solutions}
\label{results}
\subsection{Methodology}

In this work, we have prototyped all dynamical systems in Python. In particular, we did not apply any problem- or algorithm-specific optimizations and focused on using the same code base to ensure a fair comparison. For example, the momentum method is known to bring fast convergence in gradient descent learning algorithms \cite{qian:99}. Such a method can be easily incorporated to optimize algorithms like CIMs \cite{tiunov:19}, yet it might not be as useful to other algorithms like the SBM, which is based on quantum adiabatic optimization instead of gradient descent. Similarly, while the complex coupling switching is very efficient in GD algorithms to prevent the system from being trapped in local minima  \cite{Kalinin:21}, it cannot be directly applied to other algorithms using real spin variables.

Because we are integrating ODEs, it is important to choose the same integration method for all studied model systems. For the benchmark results in this section, all the algorithms are integrated using the Euler method with tunable time steps and total number of iterations. In Sec.~\ref{Method} we explain why we chose the Euler integration after a detailed study of different integration methods. Unless otherwise specified, we use Ising benchmark problems with planted solutions on a square lattice. While these problems are not NP-hard, the tunability of the planted problems \cite{perera:21} allows one to tune them to a regime where they challenge the solvers, but are not impossible to solve. It is worth noting that any results {\em strongly} depend on the benchmark problems used. However, we expect this comprehensive study to spark a more thorough benchmarking of other optimization heuristics leveraging the same code base. Clearly, real-world applications are rarely defined on planar square-lattice graphs. However, we felt that a benchmark leveraging sparse graphs was needed, in addition to the benchmarks leveraging typically complete graphs. 

The Ising test sets are tile-planted problems \cite{hamze:18,perera:20} generated by using the \texttt{Chook} \cite{perera:21} benchmark instance generator. These problems consist of square lattices of linear size $L$ with periodic boundary conditions in both directions and $N = L^2$ spins. Each lattice site is connected to four neighboring sites with a coupling strength that can only be $\pm1$ or $\pm2$. We also study the Viana-Bray \cite{viana:85} model with variable connectivity in Sec.~\ref{VB} to better understand if graph connectivity influences the algorithmic performance.

For each system size we generated $100$ random instances. Because we are working with heuristic algorithms, we use time-to-solution (TTS) to measure the performance, which is defined as
\begin{equation} \label{TTSDef}
    \text{TTS}(\gamma) = n(\gamma)\tau_\text{run},
\end{equation}
where $n(\gamma)$ is the number of necessary restarts (also referred to as samples, or shots) of the algorithm to find the ground state at least once with a desired probability $p_d$ for a given set of parameters denoted by $\gamma$. For practical purposes, we set a high confidence value of $p_d=0.99$. $\tau_\text{run}$ is the average run time of the algorithm measured in microseconds. It is straightforward to show \cite{ronnow:14a} that
\begin{equation}
    n(\gamma) = \frac{\log(1-p_d)}{\log(1-p_s(\gamma))},
\end{equation}
where $p_s(\gamma)$ gives the success probability, i.e., the chance of finding the ground state in a single run of the algorithm. Because the TTS depends on the choice of parameters $\gamma$, we carefully tune them for each algorithm to extract the true scaling. Note that, as we have multiple parameters including the pumping profile $\alpha$, Ising coupling strength $\beta$, noise profile $g_l$, time step $\Delta t$, the total number of iterations $N_t$, etc., the optimization is often multidimensional, which makes the benchmarking a relatively laborious task. For a given set of parameters $\gamma$ and a given problem instance, we run each algorithm at least $100$ times independently and compute $p_s(\gamma)$ as the percentage of the ground states found. This process is repeated for all $100$ instances to calculate the median TTS and the error bars through bootstrap sampling. Once the optimal TTS for all different problem sizes is obtained, we can study the scaling behavior of each algorithm by fitting the TTS to
\begin{equation} \label{DefineTTS}
    \text{TTS}_{\text{opt}} = 10^{a+bL}.
\end{equation}
The scaling exponent $b$ gives the performance of the algorithm in the asymptotic limit while the constant offset imposed by $a$ accounts for non-algorithmic factors and therefore is not indicative of overall performance. However, because in this study we have used the same code base and ran the code on the same hardware, in this study direct comparisons of $a$ across different SDNDSs is possible, but should be interpreted carefully. Although most of the optimizers studied here are better suited for GPU-based hardware, we performed the benchmarks using only a single core of the CPUs for uniformity.

\begin{figure}[t!]
  \centering \includegraphics[width=0.48\textwidth]{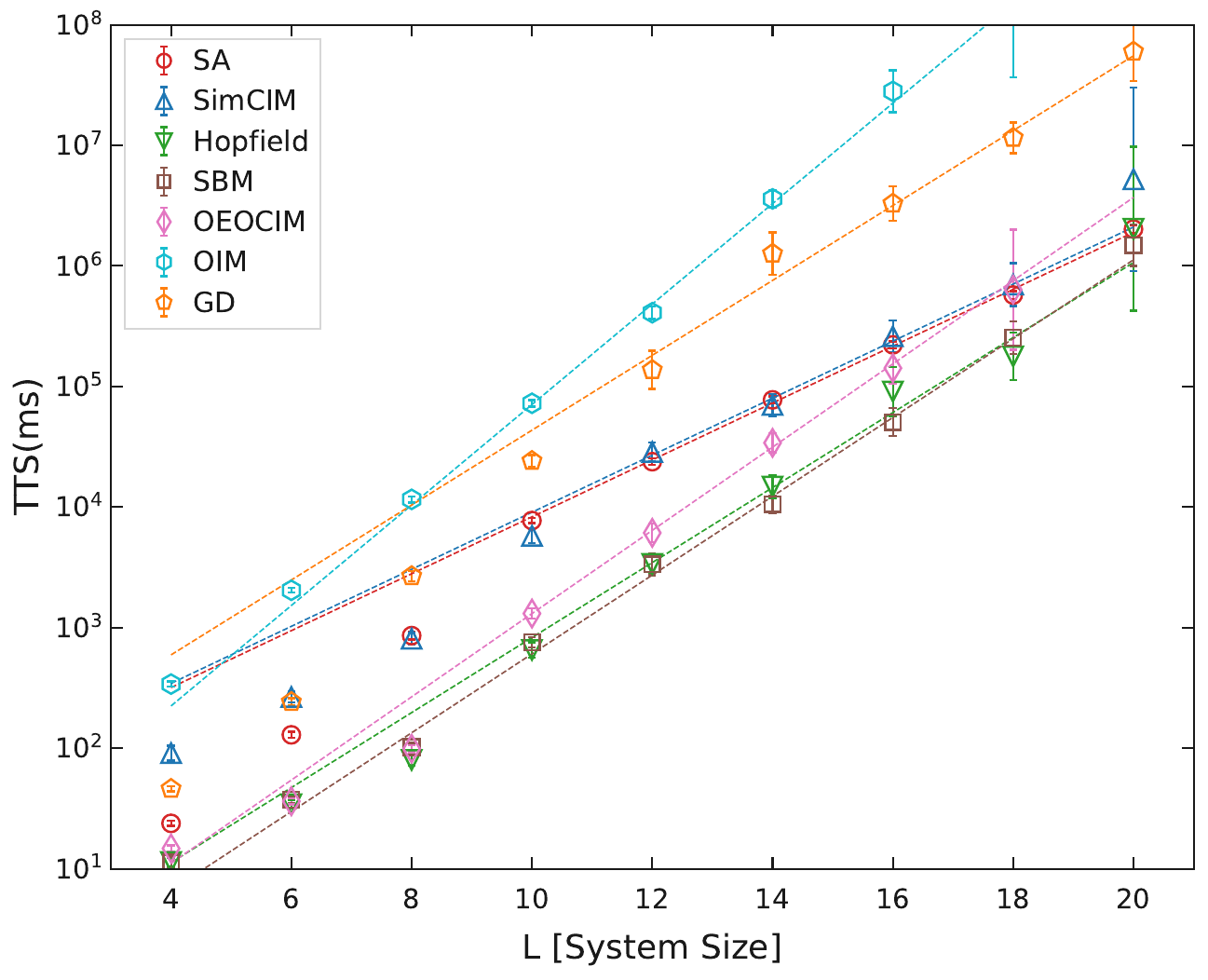}
  \caption{Time-to-solution (TTS) of various SDNDSs listed in Table~\ref{Table1: Summary of models} versus the linear size of the two-dimensional lattice. The dashed curves represent linear fits in the logarithmic (base 10) scale to the five largest system sizes.}
  \label{FigTTS}
\end{figure}

\subsection{Benchmarking results}

Figure~\ref{FigTTS} shows the scaling of various SDNDSs. Note that the exponential fits are done using the five largest problem sizes to better capture the asymptotic behavior. In Fig.~\ref{FigTTSSlope}, we compare the algorithms based on their scaling exponent $b$, that is, the slope of the corresponding linear fit (in logarithmic scale) where the height of the bars represents the uncertainty of the fit, as discussed in Sec.~\ref{slope}. 

\begin{table}[t!]
\begin{center}
\caption{
Optimal parameters for the Sim-CIM simulations for different linear problem sizes $L$. $\alpha_0$ and $\alpha_1$ determine the pumping profile (hyperbolic in this case), $\beta$ is the global Ising coupling strength, $x_\text{sat}$ is the spin saturation value, $\sigma_g$ is the width of the Gaussian noise function, and $N_t$ is the total number of iterations which is held fixed here.}
\label{TABSim-CIM}
\begin{tabular*}{\columnwidth}{@{\extracolsep{\fill}} l l l l l l l}
\hline
\hline
$L$ & $\alpha_0$ & $\alpha_1$ & $\beta$ & $x_\text{sat}$ & $\sigma_g$ & $N_t$\\
\hline
$4$  & $0.7$ & $1$ & $0.2$ & $3$  & $0.1$ & $1000$\\
$6$  & $0.7$ & $1$ & $0.2$ & $3$  & $0.1$ & $1000$\\
$8$  & $0.6$ & $1.5$ & $0.2$ & $3$  & $0.1$ & $1000$\\
$10$ & $0.6$ & $1.5$ & $0.2$ & $3$  & $0.1$ & $1000$\\
$12$ & $0.6$ & $1.5$ & $0.2$ & $3$  & $0.1$ & $1000$\\
$14$ & $0.6$ & $1.5$ & $0.2$ & $3$  & $0.1$ & $1000$\\
$16$ & $0.6$ & $1.5$ & $0.2$ & $5$  & $0.2$ & $1000$\\
$18$ & $0.6$ & $2$ & $0.2$ & $5$  & $0.2$ & $1000$\\
$20$ & $0.6$ & $2.5$ & $0.2$ & $5$  & $0.2$ & $1000$\\
\hline
\hline
\end{tabular*}
\end{center}
\end{table}

\subsubsection{Simulated coherent Ising machine}
For the Sim-CIM, there are several tunable parameters including the hyperbolic pumping profile (sharpness $\alpha_1$ and maximum pumping strength $\alpha_0$), the global Ising coupling strength $\beta$, and the spin saturation value $x_\text{sat}$. Given the problem sizes studied in this work, we use a constant total number of iterations of $N_t = 1000$. We present the optimized parameters of the Sim-CIM algorithm in Table~\ref{TABSim-CIM} and the corresponding TTS scaling in Fig.~\ref{FigTTS}. Due to the fixed number of steps, the Sim-CIM has a larger TTS for smaller problems, but it becomes competitive as the problem size increases. Our experiments suggest that the performance of the Sim-CIM is sensitive to parameter choices. Although $\alpha_0$ and $\beta$ are relatively robust, $\alpha_1$, $x_\text{sat}$, and $\sigma_g$ vary depending on the size of the problem. 

\subsubsection{Optoelectronic oscillator coherent Ising machine}
The OEO-CIM has three tunable parameters, namely, the gain strength $\alpha$, the global Ising coupling $\beta$, and the noise level $\sigma_g$. In Ref.~\cite{boehm:19}, constants $\alpha=0.25$ and $\beta=0.29$ are chosen for square lattice problems. These values are consistent with our optimized results reported in Table~\ref{TABOEO-CIM}. Because the OEO-CIM converges relatively fast in comparison to the Sim-CIM---usually in a few hundred steps---it is very competitive for small system sizes. However, the algorithm becomes unstable with larger system sizes, leading to unreasonably large errors in the TTS when $L>10$. Thus, the benchmark results for the OEO-CIM in Fig.~\ref{FigTTS} are obtained by introducing a time-dependent gain profile. Table~\ref{TABOEO-CIM} shows that the optimal parameters of the OEO-CIM follow a trend similar to that of the parameters of the Sim-CIM. However, we observe a sudden jump in the gain sharpness $\alpha_1$ at $L=18$, beyond which the OEO-CIM can rarely find the ground state after $100$ restarts for any of the $100$ instances studied.

\begin{table}[h]
\begin{center}
\caption{
Optimal parameters for the OEO-CIM simulations for different linear problem sizes $L$. $\alpha_0$ and $\alpha_1$ define a hyperbolic pumping profile, $\beta$ is the global Ising coupling strength, $\sigma_g$ is the Gaussian noise level, and $N_t$ is the total number of iterations. }
\label{TABOEO-CIM}
\begin{tabular*}{\columnwidth}{@{\extracolsep{\fill}} l l l l l l}
\hline
\hline
$L$ & $\alpha_0$ & $\alpha_1$ & $\beta$ & $\sigma_g$ & $N_t$\\
\hline
$4$  & $0.45$ & $2$ & $0.29$ & $0.1$ & $2^6$\\
$6$  & $0.45$ & $3.5$ & $0.29$ & $0.1$ & $2^6$\\
$8$  & $0.45$ & $5$ & $0.29$ & $0.1$ & $2^7$\\
$10$  & $0.45$ & $5$ & $0.29$ & $0.1$ & $2^7$\\
$12$  & $0.45$ & $5$ & $0.29$ & $0.2$ & $2^8$\\
$14$  & $0.45$ & $5$ & $0.29$ & $0.2$ & $2^8$\\
$16$  & $0.45$ & $5$ & $0.29$ & $0.2$ & $2^9$\\
$18$  & $0.45$ & $9$ & $0.29$ & $0.2$ & $2^9$\\
\hline
\hline
\end{tabular*}
\end{center}
\end{table}

\subsubsection{Gain dissipative algorithm}
We choose the parameters of the GD algorithm following the original proposal of Ref.~\cite{kalinin:18a}. Specifically, we use a uniform local Ising coupling $\beta_{lm}=\beta=1$ and $K_{lm}=J_{lm}$ such that only two of three equations of motion are nontrivial (see Table~\ref{Table1: Summary of models}). The remaining parameters are chosen as
\begin{eqnarray}
    &&\epsilon=0.005\,\max_l\sum_m|J_{lm}|,\\
    &&\rho_\text{th}=0.15\, \max_l\sum_m|J_{lm}|,\\
    &&\gamma(t=0) = -\max_l\sum_m|J_{lm}|,    
\end{eqnarray}
and $\Psi_i(t=0)=1$. Finally, $h_{2m}$ has a hyperbolic tangent profile specified by $\alpha_1$ and $\alpha_0$.
In our experiments, we find that, unlike CIMs, GD systems can be robust in terms of parameter choices, which is consistent with the authors' results on Max-Cut problems reported in Ref.~\cite{kalinin:18a}. Thus, we have fixed the gain profile and the global Ising coupling for different problem sizes and only tuned the total number of iterations. The parameters used in the simulations are summarized in Table~\ref{TABGD}. Figure~\ref{FigTTS} shows that our implementation of the GD algorithm has a relatively large TTS for all problem sizes studied. We conjecture that the synchronization of the network may take extra time due to the small pumping rate in our experiments and because numerical computations using complex numbers are slower than when using real numbers.

\begin{table}[h]
\begin{center}
\caption{
Total number of iterations $N_t$ for different dynamical system algorithms versus linear problem sizes $L$ used in the simulations. For simulated annealing (SA), we use an inverse linear annealing schedule and set the final inverse temperature to $\beta_f=3$. For the GD system we fix $\alpha_0=0.5$, $\alpha_1=10$, $\beta=1$, and $\sigma_g=0.2$. For the SBM, the time step is fixed to $dt=0.5$. In the case of the Hopfield model dynamical system, we set $\zeta_0=10$, $\zeta_d=0.8$, $dt=0.5$, and $\sigma_g=0.1$. For the HNN model with fourth-order Runge-Kutta (RK4) and Runge-Kutta-Fehlberg (RK45) integration schemes, we fix the initial time step to $dt=1$ and $dt_0=1$, respectively. We also choose $\epsilon=1$ to curb errors.}
\label{TABGD}
\begin{tabular*}{\columnwidth}{@{\extracolsep{\fill}}lllllllllll}
\hline
\hline
\multicolumn{1}{l}{$L$} & $4$ & $6$ & $8$ & $10$ & $12$ & $14$ & $16$ & $18$ & $20$\\
\hline
SA  & $2^2$ & $2^4$ & $2^4$ & $2^9$ & $2^9$ & $2^{10}$ & $2^{10}$ & $2^{13}$ & $2^{13}$\\
GD  & $2^6$ & $2^7$ & $2^8$ & $2^9$ & $2^{10}$ & $2^{10}$ & $2^{12}$ & $2^{14}$ & $2^{14}$\\
SBM  & $2^5$ & $2^7$ & $2^7$ & $2^8$ & $2^8$ & $2^8$ & $2^9$ & $2^9$ & $2^{10}$\\
Hopfield  & $2^5$ & $2^7$ & $2^7$ & $2^8$ & $2^8$ & $2^8$ & $2^9$ & $2^9$ & $2^{10}$\\
HNN(RK4) & $2^5$ & $2^5$ & $2^6$ & $2^6$ & $2^6$ & $2^7$ & $2^8$ & -- & -- &\\
HNN(RK45) & $2^8$ & $2^9$ & $2^{10}$ & $2^{10}$ & $2^{10}$ & $2^{10}$ & $2^{11}$ & -- & -- &\\
\hline
\hline
\end{tabular*}
\end{center}
\end{table}

\subsubsection{Oscillator-based Ising machine}
In our experiments we follow Fig.~10 in Ref.~\cite{wang:19a} and use the parameters used in the aforementioned reference. Tests on small system sizes confirm that the choice  of parameters for the OIM is robust for different system sizes. For the OIM the gain is periodic, which helps the system escape local minima by imposing and relaxing the binary constraint repeatedly. While this approach remains untested for other dynamical system, it could potentially improve performance across the different approaches studied in this work. Figure~\ref{FigTTS} shows that the OIM has the largest TTS values of all method studied for almost all problem sizes studied.
We have set the evolution time of the OIM to $20$ steps while the optimal time step is found to be $0.02$, independent of problem size $L$. Thus, the OIM has the same total number of iterations as the Sim-CIM. This indicates that OIM has a lower success rate to find the optimal than, e.g., the Sim-CIM, and we observe larger error bars for $L=18$. It is also possible that the adapted annealing scheme might not be optimal and a more careful tuning of the parameters might improve the TTS.

\subsubsection{Simulated bifurcation machine}
For the SBM the gain profile $\alpha$ is a linear function passing through the origin. The other parameters are chosen as follows. $K=\Delta=1$ and $\beta=0.7/\sigma(J)N_s$, where $\sigma(\bullet)$ denotes the standard deviation \cite{goto:19}. As pointed out before, the SBM is different from other SDNDSs because it has a second-order equation of motion (Hamiltonian dynamics). Therefore, instead of using the Euler method, which can cause extra numerical errors in solving higher-order ODEs, we apply the symplectic Euler method \cite{goto:19}. Finally, there is no noise term in the SBM, given its origin stemming from adiabatic quantum optimization. In our experiments the parameters of SBM are mostly constant across different system sizes and we only optimize the total number of iterations listed in Table~\ref{TABGD}. From Fig.~\ref{FigTTS}, we observe that the SBM seems to be an effective optimizer for the problems studied, because overall it has one of the smallest TTS for the different system sizes. However, we note that the SBM may require much smaller time steps in some cases.

\subsubsection{Hopfield neural network dynamical system}
For the HNN dynamical system we set parameters $\alpha=\beta=1$ and $\phi(x_l)=\tanh(x_l/\zeta(t))$ \cite{joya:02}. Specifically, $\zeta(t)>0$ is a geometric function with initial value $\zeta(0)=\zeta_0$ and decay rate $\zeta_d$. Parameters do not tend to vary across system sizes studied and thus we fix these for all $L$. Remaining parameters are listed in Table~\ref{TABGD}. Note that we simulate a stochastic Hopfield model including a noise term (discrete or continuous), because it has been shown to have better performance than its deterministic counterpart \cite{cai:20,liu:09}. However, in this work we find that such positive effect is only visible for smaller system sizes. 
Our results show that the HNN dynamical system tends to become trapped in local minima. However, it remains competitive for small to medium system sizes. In terms of the absolute TTS, it is faster than most of the dynamical systems we tested (see Fig.~\ref{FigTTS}) due to its simple dynamics and fast convergence. However, when $L \geq 24$ the HNN dynamical system has difficulties in finding the optimal of the problem studied. 

\subsubsection{Simulated annealing}
As a benchmark, we also use simulated annealing. In our implementation we use an inverse linear annealing schedule and tune both the decay rate and the total number of iterations to optimal values to determine the TTS. In our experiments we find the final inverse temperature $\beta_f=3$ is independent of linear lattice size $L$. The total number of iterations for different system sizes are summarized in Table~\ref{TABGD}. As shown in Fig.~\ref{FigTTS}, the TTS of SA is comparable to the SDNDSs.

However, to better compare the asymptotic behavior of the different optimizers studied, we extract the scaling exponent (slope) of the data in Fig.~\ref{FigTTS} in Sec.~\ref{slope}.

\begin{figure}[t!]
  \centering \includegraphics[width=0.48\textwidth]{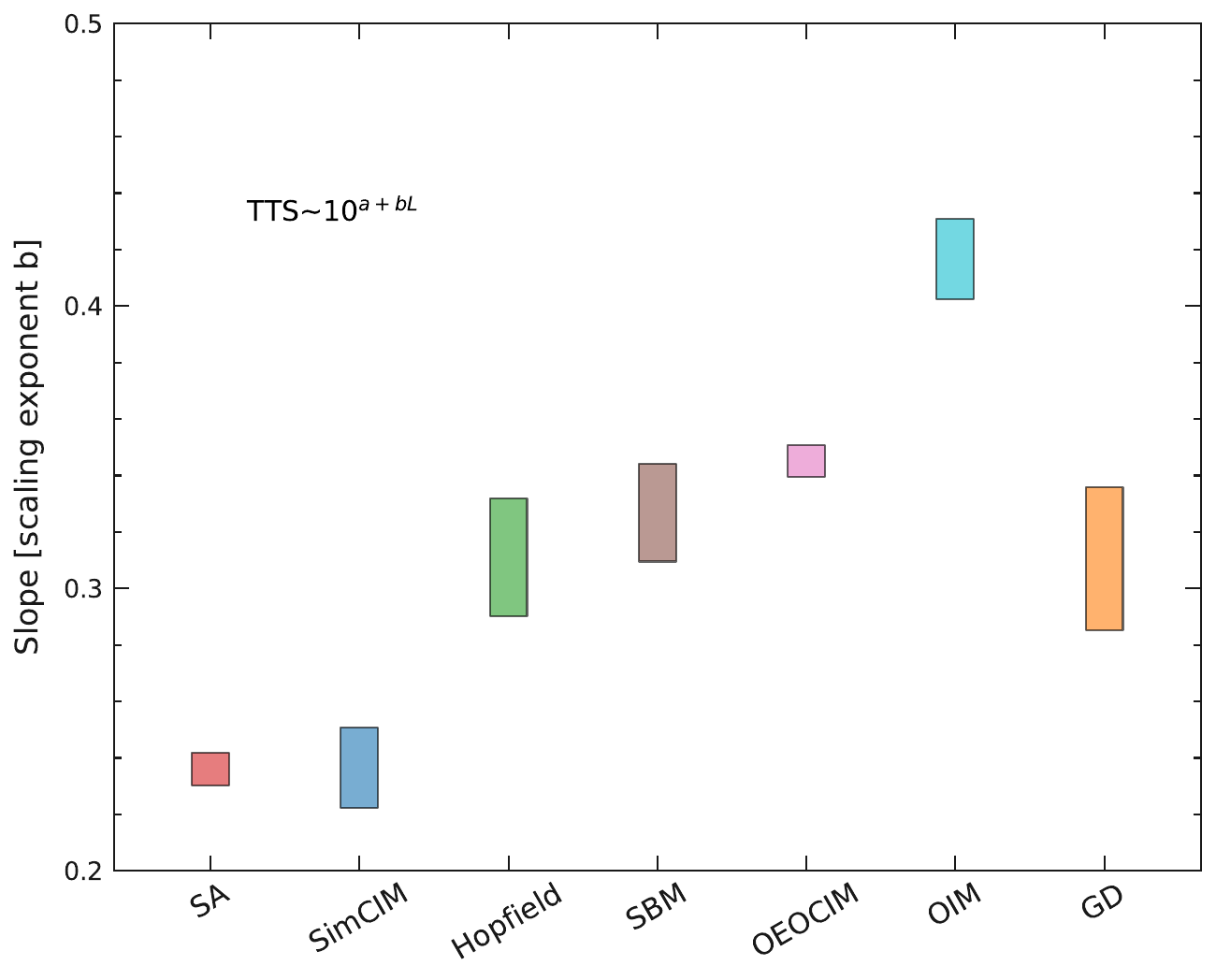}
  \caption{Scaling exponent $b$ of different SDNDSs listed in in Table~\ref{Table1: Summary of models} which corresponds to the slope of the linear fit (in logarithmic scale) to the TTS data shown in Fig.~\ref{FigTTS}. The height of the boxes represents the error bars.}
  \label{FigTTSSlope}
\end{figure}

\subsection{Scaling exponent}
\label{slope}

Figure \ref{FigTTSSlope} shows the scaling exponent $b$ for the different optimization methods studied, extracted from the data in Fig.~\ref{FigTTS} by fitting a linear function to the $5$ largest system sizes in a log-linear sale. Interestingly, SA and the SimCIM have comparable asymptotic performance and show the best scaling for the benchmark problems and system sizes studied. This corroborates  claims that some of the SDNDS algorithms can be as efficient as some modern heuristics \cite{leleu:19}. Additional experiments for SA (not shown) suggest that the scaling exponent slightly decreases further for increasing system sizes. However, to perform an apples-to-apples comparison we decided to restrict the analysis to $L\leq 24$. The main reason for this is that many of the SDNDSs are unable to find the optima for large system sizes, even after hundreds of iterations. We performed experiments for larger system sizes (not shown) measuring the time-to-target, but in that case most problems were too easy to solve and thus no meaningful results could be extracted.

All other dynamical systems, except the OIM discussed below, have comparable scaling exponents (within error bars), which can be ascribed to the similarities between their equations of motion and corresponding dynamics, see Table~\ref{Table1: Summary of models}. We suspect that the better performance of the Sim-CIM for the problems studied can be ascribed to both larger time steps than other SDNDSs, as well as its unique nonlinear activation, which has also improved the performance of the SBM \cite{goto:21}. However, the improved performance of the Sim-CIM comes at the price of tuning more parameters. Finally, note that the suboptimal performance of the OIM might be due to the fact parameter tuning for this SDNDS is difficult.

\section{Additional experiments}
\label{additional_ex}

In Sec.~\ref{results} we discussed the benchmark results of different SDNDSs on small- to medium-sized random Ising problems and compared performance to SA. In this section we analyze the choice of integrator and the effects of lattice connectivity.

\subsection{Effects of numerical integration methods} 
\label{Method}

Algorithms like Sim-CIM, OEO-CIM, and the discrete-time Hopfield neural network are discrete in time. This means that there is no need to specify a numerical integration method. However, algorithms such as OPO-CIM, GD, and OIM, are first-order ordinary differential equations (ODEs). While the Euler method is the most common choice to integrate these, few efforts have been devoted to investigating how numerical methods affect these optimization routines.

A special case is the SBM which is a second-order ODE following a semi-classical Hamiltonian formalism. Thus, we can solve the ODE leveraging a symplectic Euler method. Note that it has been shown that a symplectic Euler method can accelerate the algorithm by reducing the required time step by half \cite{goto:19}. Furthermore, we empirically determined that using the St\"{o}rmer-Verlet method does not improve the scaling.

As an example, we compare the Euler method to higher-order methods such as the fourth-order RK method (RK4) and the Runge-Kutta-Fehlberg method (RK45). Note that, the latter has adaptive step sizes. In general, a higher-order method indicates higher accuracy. In other words, the algorithm may have larger time steps, making it possible to find the solutions within fewer iterations. However, these methods usually require more computational steps during each iteration. For RK4 and RK45, the ODEs are evaluated three and five more times, respectively, than the simple Euler method during each iteration. 

\begin{figure}[h]
  \centering \includegraphics[width=0.48\textwidth]{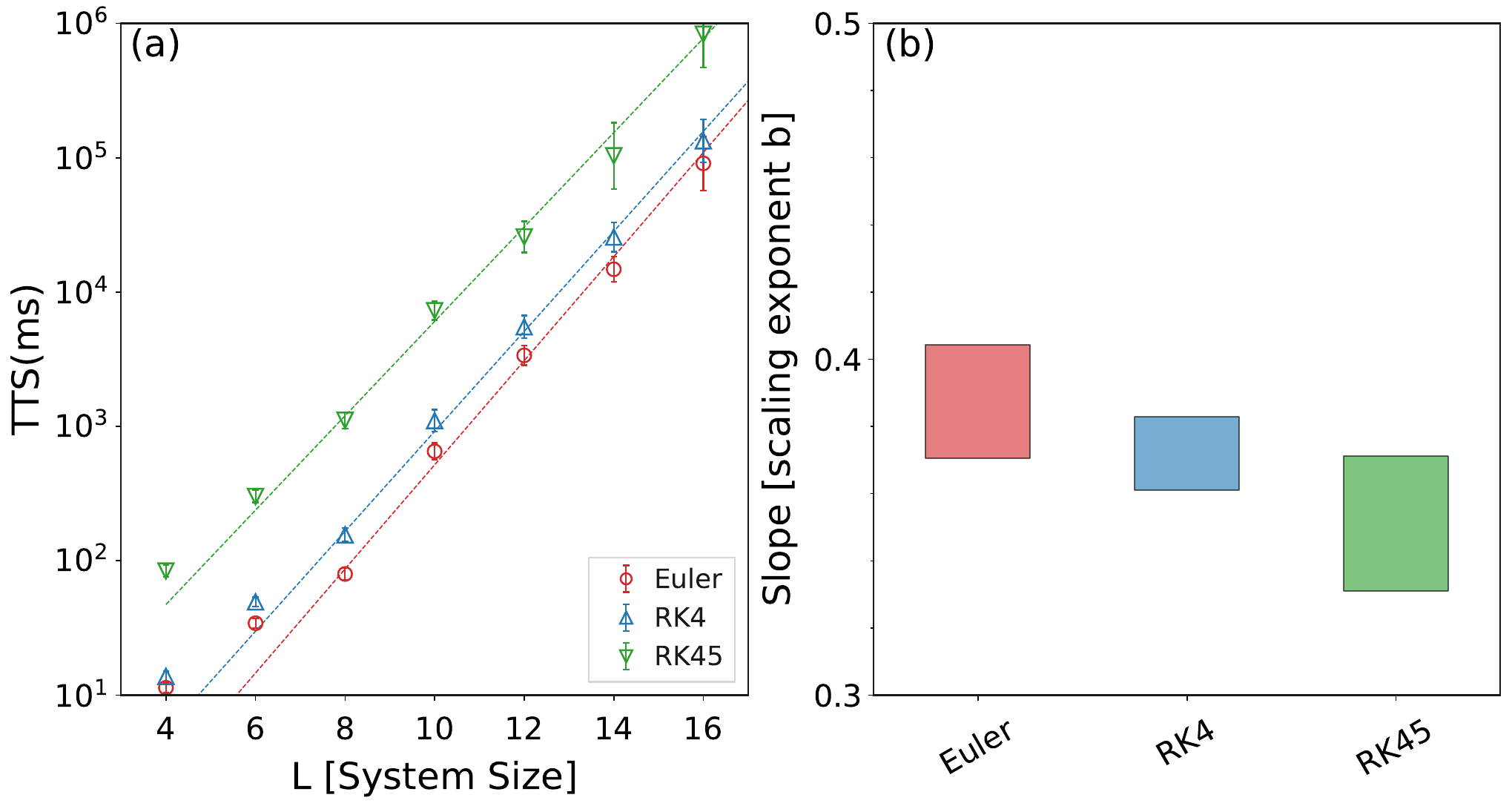}
  \caption{Comparison between the scaling results of different integration methods tested on the continuous Hopfield neural network model. (a) Time-to-solution (TTS) of three numerical methods in a logarithmic (base 10) scale versus the linear lattice size. The dashed curves represent linear fits to the five largest system sizes. (b) The scaling exponent $b$, corresponding to the data in panel (a). The height of the boxes represents the error bars. Note that we cut the data at $L=16$ so that the slope of the Euler method is slightly larger than that in Fig.~\ref{FigTTSSlope}. The scaling exponent agrees within error bars for the different integration schemes used. }
  \label{FigMethod}
\end{figure}

Figure \ref{FigMethod}(a) shows the scaling results of the Hopfield neural network with different integration methods. The parameters for the Euler method are the same as in Table~\ref{TABGD}. Note that, for RK4, we find that a larger step size ($dt=1$) also works well. For both higher-order integrators (RK4 and RK45) the overall numerical effort is larger than for the Euler method. Given that the scaling exponents agree within error bars, there is no clear indication that using higher-order schemes is needed when integrating the equations of motion. 

\begin{figure}[h]
  \centering \includegraphics[width=0.48\textwidth]{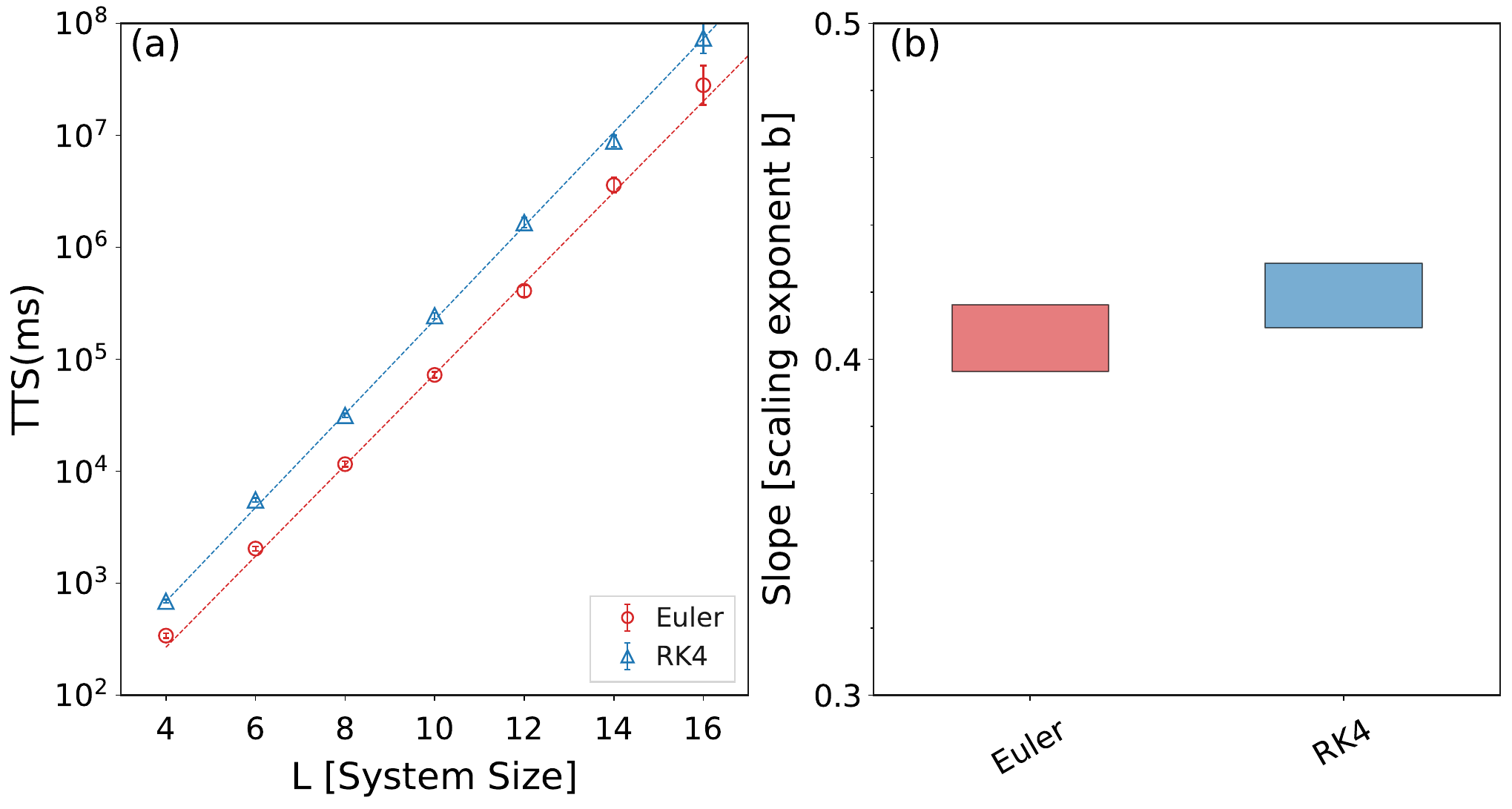}
  \caption{Comparison of different integration methods tested on the OIM. (a) Time-to-solution (TTS) of two numerical integration methods in a logarithmic (base 10) scale versus the linear size $L$ of the problem. The dashed curves represent linear fits to the five largest system sizes. (b) The scaling exponent $b$, corresponding to the data in panel (a). The height of the boxes represents the error bars.}
  \label{FigMethodOIM}
\end{figure}

We further test RK4 for the OIM, because it has relatively complicated dynamics with periodic gains. A comparison between the Euler and RK4 methods is shown in Fig.~\ref{FigMethodOIM}. For $L = 4$, $6$, $8$, and $10$ we use $dt = 0.05$, whereas for $L \geq 12$ we use $dt = 0.02$. Recall that the optimal time step for OIM using the Euler integration scheme is $dt = 0.02$, which is independent of system size. Thus, for small problems $L\leq10$, RK4 can bear a larger time step $dt=0.05$. Again, as can be seen in Fig.~\ref{FigMethodOIM} the scaling exponent $b$ agrees within error bars and a more sophisticated integrator means only additional numerical overhead. 

Summarizing, using different integration schemes does not seem to improve the TTS scaling for both the Hopfield neural network and the OIM. This is in agreement with results for digital memcomputing machines \cite{zhang:21} based on memristor circuits. One potential use for RK45 could be for initialization. After a short, burn-in phase an optimal time step is determined, after which the integrator is swapped out by a simple Euler method. Such approach would drastically simplify parameter tuning.

\subsection{Effects of graph connectivity} \label{VB}

So far, we have focused on two-dimensional Ising models with a fixed connectivity of 4 between the variables. Considering the collective dynamics of SDNDSs, it is reasonable to suspect that performance might be affected by the average connectivity of the graph. In fact, recent studies \cite{andresen:13} suggest that large system-spanning rearrangements of variables---needed to effectively traverse the energy landscape---can only occur when graph connectivity is high. Furthermore, full self-organized criticality---an effect where there is no characteristic size of a variable rearrangement---only sets in when graphs are fully connected. This suggests that the performance of SDNDSs should improve, the denser the graph is. 

\begin{figure}[h]
  \centering \includegraphics[width=0.48\textwidth]{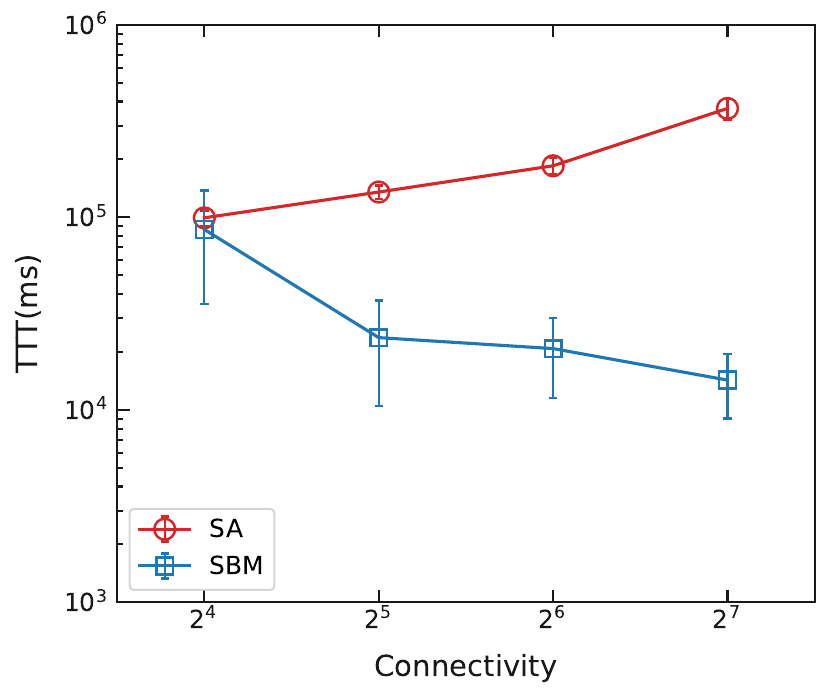}
  \caption{Time-to-solution (TTS) in a logarithmic (base 10) scale versus the connectivity for the Viana-Bray (VB) model with $2^7$ variables with Gaussian couplers between the spins. The figure compares the TTS for SA and SBM when the connectivity between the variables is increased in powers of 2 until the graph is complete. While for SA the TTS increases with increasing graph density due to the fact that the number of Hamiltonian terms increases, for the SBM the TTS decreases. This suggests that the higher the connectivity, the more efficient SDNDSs such as the SBM are at finding optima of cost functions. For SA the final inverse temperature is $\beta_f=3$ and the total number of sweeps is $N_t=2^{10}$. For the SBM we use a time step of $dt=0.25$ and $N_t=2^{11}$ total algorithmic steps.}
  \label{FigVB}
\end{figure}

To test this hypothesis we use the Viana-Bray (VB) model \cite{viana:85}. For example, for the Viana-Bray model with connectivity 3, each spin is connected to three other randomly chosen spins, i.e., the system has no geometry despite a fixed connectivity. We study a small VB model with $N = 2^7 = 128$ variables and successively increase the connectivity in powers of $2$ from $16$ to $128$, where the latter represents the fully-connected Sherrington-Kirkpatrick spin-glass model \cite{sherrington:75}. We use both SA and the SBM to to find optima for this system as a function of connectivity. Data are averaged over 100 disorder realizations of Gaussian disorder with zero mean and standard deviation unity. The results are summarized in Fig.~\ref{FigVB}. The data clearly show that the TTS increases with connectivity for SA. This is no surprise, because the number of Hamiltonian terms grows with the connectivity and thus the time it takes for one Monte Carlo sweep. In contrast, the TTS decreases for increasing connectivity for the SBM. This suggests that large avalanches are needed for SDNDSs to be effective optimizers. This, in turn, suggests that the denser the graph of the problem to be optimized, the more effective SDNDSs are, and that self-organized critical behavior resulting in large spin rearrangements is critical for optimization in dynamical systems. 

\section{Discussions and conclusions}

In this work we have examined several representative combinatorial Ising solvers based on stochastic dynamical nonlinear systems---where binary variables are relaxed to continuous analog spins---and compared them using the same software stack optimizing square-lattice disordered Ising spin-glass problems with planted solutions. A list of the different equations of motion is given in Table~\ref{Table1: Summary of models}. All dynamical systems share some characteristics, namely a driver term $\alpha$, a global Ising coupling $\beta$, a noise term $g_l$, and nonlinear dynamics. In addition, we compared the performance of the driven dynamical systems to simulated annealing as a benchmark. Our results illustrate the advantages and disadvantages some of these stochastic-driven nonlinear dynamical systems pose (in comparison to simulated annealing) for a given benchmark problem and moderate system sizes. It would be interesting to do further comparisons for larger system sizes, still leveraging the same software stack. However, this was outside the scope of this project and would require implementations leveraging GPUs. We also analyze the effects of higher-order integration schemes and show that simple Euler dynamics seems, in general, sufficient for most applications. 

We also analyzed the effects of graph connectivity on the solver performance. Optimizers require large variable rearrangements to be effective when traversing the search space. Our conjecture that self-organized critical behavior is needed for SDNDS solvers to be effective is backed by numerical results that show that when the connectivity is increased, performance increases despite the number of terms to sum over in the cost function increases as well. In general, most optimization problems are not natively QUBO and also often have constraints that need to be added to the cost function. In the process of mapping optimization problems to QUBOs, as well as including constraints via quadratization into the cost function, the resulting graph tends to be highly if not fully connected. This means that optimization heuristics based on stochastic-driven nonlinear dynamical systems have an inherent advantage over annealing-based quantum devices \cite{lanting:14} for real-world applications.

Finally, the dynamical systems listed in Table~\ref{Table1: Summary of models} stem from physical phenomena. It would be interesting to investigate if better equations of motion that do not follow physical processes can be found with the goal of improving performance. 

\begin{acknowledgments}
We would like to thank Ken Zick for useful comments.
\end{acknowledgments}

\bibliographystyle{apsrevtitle}
\bibliography{refs}

\begin{thebibliography}{64}
\expandafter\ifx\csname natexlab\endcsname\relax\def\natexlab#1{#1}\fi
\expandafter\ifx\csname bibnamefont\endcsname\relax
  \def\bibnamefont#1{#1}\fi
\expandafter\ifx\csname bibfnamefont\endcsname\relax
  \def\bibfnamefont#1{#1}\fi
\expandafter\ifx\csname citenamefont\endcsname\relax
  \def\citenamefont#1{#1}\fi
\expandafter\ifx\csname url\endcsname\relax
  \def\url#1{\texttt{#1}}\fi
\expandafter\ifx\csname urlprefix\endcsname\relax\def\urlprefix{URL }\fi
\providecommand{\bibinfo}[2]{#2}
\providecommand{\eprint}[2][]{\url{#2}}

\bibitem[{\citenamefont{Arora and Barak}(2009)}]{arora:09}
\bibinfo{author}{\bibfnamefont{S.}~\bibnamefont{Arora}} \bibnamefont{and}
  \bibinfo{author}{\bibfnamefont{B.}~\bibnamefont{Barak}},
  \emph{\bibinfo{title}{Computational complexity: a modern approach}}
  (\bibinfo{publisher}{Cambridge University Press},
  \bibinfo{address}{Cambridge, U.K.}, \bibinfo{year}{2009}).

\bibitem[{\citenamefont{Smith}(1999)}]{smith:99}
\bibinfo{author}{\bibfnamefont{K.~A.} \bibnamefont{Smith}},
  \emph{\bibinfo{title}{Neural networks for combinatorial optimization: A
  review of more than a decade of research}}, \bibinfo{journal}{INFORMS Journal
  on Computing} \textbf{\bibinfo{volume}{11}}, \bibinfo{pages}{15}
  (\bibinfo{year}{1999}).

\bibitem[{\citenamefont{Juan et~al.}(2015)\citenamefont{Juan, Faulin, Grasman,
  Rabe, and Figueira}}]{juan:15}
\bibinfo{author}{\bibfnamefont{A.~A.} \bibnamefont{Juan}},
  \bibinfo{author}{\bibfnamefont{J.}~\bibnamefont{Faulin}},
  \bibinfo{author}{\bibfnamefont{S.~E.} \bibnamefont{Grasman}},
  \bibinfo{author}{\bibfnamefont{M.}~\bibnamefont{Rabe}}, \bibnamefont{and}
  \bibinfo{author}{\bibfnamefont{G.}~\bibnamefont{Figueira}},
  \emph{\bibinfo{title}{A review of simheuristics: Extending metaheuristics to
  deal with stochastic combinatorial optimization problems}},
  \bibinfo{journal}{Operations Research Perspectives}
  \textbf{\bibinfo{volume}{2}}, \bibinfo{pages}{62} (\bibinfo{year}{2015}),
  ISSN \bibinfo{issn}{2214-7160}.

\bibitem[{\citenamefont{Barahona}(1982)}]{barahona:82}
\bibinfo{author}{\bibfnamefont{F.}~\bibnamefont{Barahona}},
  \emph{\bibinfo{title}{On the computational complexity of ising spin glass
  models}}, \bibinfo{journal}{Journal of Physics A: Mathematical and General}
  \textbf{\bibinfo{volume}{15}}, \bibinfo{pages}{3241} (\bibinfo{year}{1982}).

\bibitem[{\citenamefont{Lucas}(2014)}]{lucas:14}
\bibinfo{author}{\bibfnamefont{A.}~\bibnamefont{Lucas}},
  \emph{\bibinfo{title}{{Ising formulations of many NP problems}}},
  \bibinfo{journal}{Front. Physics} \textbf{\bibinfo{volume}{12}},
  \bibinfo{pages}{5} (\bibinfo{year}{2014}).

\bibitem[{\citenamefont{Kirkpatrick et~al.}(1983)\citenamefont{Kirkpatrick,
  {Gelatt, Jr.}, and Vecchi}}]{kirkpatrick:83}
\bibinfo{author}{\bibfnamefont{S.}~\bibnamefont{Kirkpatrick}},
  \bibinfo{author}{\bibfnamefont{C.~D.} \bibnamefont{{Gelatt, Jr.}}},
  \bibnamefont{and} \bibinfo{author}{\bibfnamefont{M.~P.}
  \bibnamefont{Vecchi}}, \emph{\bibinfo{title}{Optimization by simulated
  annealing}}, \bibinfo{journal}{Science} \textbf{\bibinfo{volume}{220}},
  \bibinfo{pages}{671} (\bibinfo{year}{1983}).

\bibitem[{\citenamefont{{Boettcher} and {Percus}}(2001)}]{boettcher:01}
\bibinfo{author}{\bibfnamefont{S.}~\bibnamefont{{Boettcher}}} \bibnamefont{and}
  \bibinfo{author}{\bibfnamefont{A.~G.} \bibnamefont{{Percus}}},
  \emph{\bibinfo{title}{{{Optimization with Extremal Dynamics}}}},
  \bibinfo{journal}{Phys. Rev. Lett.} \textbf{\bibinfo{volume}{86}},
  \bibinfo{pages}{5211} (\bibinfo{year}{2001}).

\bibitem[{\citenamefont{Wang et~al.}(2015)\citenamefont{Wang, Machta, and
  Katzgraber}}]{wang:15c}
\bibinfo{author}{\bibfnamefont{W.}~\bibnamefont{Wang}},
  \bibinfo{author}{\bibfnamefont{J.}~\bibnamefont{Machta}}, \bibnamefont{and}
  \bibinfo{author}{\bibfnamefont{H.~G.} \bibnamefont{Katzgraber}},
  \emph{\bibinfo{title}{{Comparing Monte Carlo methods for finding ground
  states of Ising spin glasses: Population annealing, simulated annealing, and
  parallel tempering}}}, \bibinfo{journal}{Phys. Rev. E}
  \textbf{\bibinfo{volume}{92}}, \bibinfo{pages}{013303}
  (\bibinfo{year}{2015}).

\bibitem[{\citenamefont{Sutton et~al.}(2017)\citenamefont{Sutton, Camsari,
  Behin-Aein, and Datta}}]{sutton:17}
\bibinfo{author}{\bibfnamefont{B.}~\bibnamefont{Sutton}},
  \bibinfo{author}{\bibfnamefont{K.~Y.} \bibnamefont{Camsari}},
  \bibinfo{author}{\bibfnamefont{B.}~\bibnamefont{Behin-Aein}},
  \bibnamefont{and} \bibinfo{author}{\bibfnamefont{S.}~\bibnamefont{Datta}},
  \emph{\bibinfo{title}{Intrinsic optimization using stochastic nanomagnets}},
  \bibinfo{journal}{Scientific Reports} \textbf{\bibinfo{volume}{7}}
  (\bibinfo{year}{2017}), ISSN \bibinfo{issn}{2045-2322}.

\bibitem[{\citenamefont{Kadowaki and Nishimori}(1998)}]{kadowaki:98}
\bibinfo{author}{\bibfnamefont{T.}~\bibnamefont{Kadowaki}} \bibnamefont{and}
  \bibinfo{author}{\bibfnamefont{H.}~\bibnamefont{Nishimori}},
  \emph{\bibinfo{title}{{{Quantum annealing in the transverse Ising model}}}},
  \bibinfo{journal}{Phys. Rev. E} \textbf{\bibinfo{volume}{58}},
  \bibinfo{pages}{5355} (\bibinfo{year}{1998}).

\bibitem[{\citenamefont{{Farhi} et~al.}(2000)\citenamefont{{Farhi},
  {Goldstone}, {Gutmann}, and {Sipser}}}]{farhi:00}
\bibinfo{author}{\bibfnamefont{E.}~\bibnamefont{{Farhi}}},
  \bibinfo{author}{\bibfnamefont{J.}~\bibnamefont{{Goldstone}}},
  \bibinfo{author}{\bibfnamefont{S.}~\bibnamefont{{Gutmann}}},
  \bibnamefont{and} \bibinfo{author}{\bibfnamefont{M.}~\bibnamefont{{Sipser}}},
  \emph{\bibinfo{title}{{{Quantum Computation by Adiabatic Evolution}}}}
  (\bibinfo{year}{2000}), \bibinfo{note}{arXiv:quant-ph/0001106}.

\bibitem[{\citenamefont{Leleu et~al.}(2019)\citenamefont{Leleu, Yamamoto,
  McMahon, and Aihara}}]{leleu:19}
\bibinfo{author}{\bibfnamefont{T.}~\bibnamefont{Leleu}},
  \bibinfo{author}{\bibfnamefont{Y.}~\bibnamefont{Yamamoto}},
  \bibinfo{author}{\bibfnamefont{P.~L.} \bibnamefont{McMahon}},
  \bibnamefont{and} \bibinfo{author}{\bibfnamefont{K.}~\bibnamefont{Aihara}},
  \emph{\bibinfo{title}{Destabilization of local minima in analog spin systems
  by correction of amplitude heterogeneity}}, \bibinfo{journal}{Phys. Rev.
  Lett.} \textbf{\bibinfo{volume}{122}}, \bibinfo{pages}{040607}
  (\bibinfo{year}{2019}).

\bibitem[{\citenamefont{Wang et~al.}(2013)\citenamefont{Wang, Marandi, Wen,
  Byer, and Yamamoto}}]{wang:13b}
\bibinfo{author}{\bibfnamefont{Z.}~\bibnamefont{Wang}},
  \bibinfo{author}{\bibfnamefont{A.}~\bibnamefont{Marandi}},
  \bibinfo{author}{\bibfnamefont{K.}~\bibnamefont{Wen}},
  \bibinfo{author}{\bibfnamefont{R.~L.} \bibnamefont{Byer}}, \bibnamefont{and}
  \bibinfo{author}{\bibfnamefont{Y.}~\bibnamefont{Yamamoto}},
  \emph{\bibinfo{title}{{Coherent Ising machine based on degenerate optical
  parametric oscillators}}}, \bibinfo{journal}{Phys. Rev. A}
  \textbf{\bibinfo{volume}{88}}, \bibinfo{pages}{063853}
  (\bibinfo{year}{2013}).

\bibitem[{\citenamefont{Marandi et~al.}(2014)\citenamefont{Marandi, Wang,
  Takata, Byer, and Yamamoto}}]{marandi:14}
\bibinfo{author}{\bibfnamefont{A.}~\bibnamefont{Marandi}},
  \bibinfo{author}{\bibfnamefont{Z.}~\bibnamefont{Wang}},
  \bibinfo{author}{\bibfnamefont{K.}~\bibnamefont{Takata}},
  \bibinfo{author}{\bibfnamefont{R.~L.} \bibnamefont{Byer}}, \bibnamefont{and}
  \bibinfo{author}{\bibfnamefont{Y.}~\bibnamefont{Yamamoto}},
  \emph{\bibinfo{title}{Network of time-multiplexed optical parametric
  oscillators as a coherent ising machine}}, \bibinfo{journal}{Nature
  Photonics} \textbf{\bibinfo{volume}{8}}, \bibinfo{pages}{937–942}
  (\bibinfo{year}{2014}), ISSN \bibinfo{issn}{1749-4893}.

\bibitem[{\citenamefont{Inagaki
  et~al.}(2016{\natexlab{a}})\citenamefont{Inagaki, Inaba, Hamerly, Inoue,
  Yamamoto, and Takesue}}]{inagaki:16a}
\bibinfo{author}{\bibfnamefont{T.}~\bibnamefont{Inagaki}},
  \bibinfo{author}{\bibfnamefont{K.}~\bibnamefont{Inaba}},
  \bibinfo{author}{\bibfnamefont{R.}~\bibnamefont{Hamerly}},
  \bibinfo{author}{\bibfnamefont{K.}~\bibnamefont{Inoue}},
  \bibinfo{author}{\bibfnamefont{Y.}~\bibnamefont{Yamamoto}}, \bibnamefont{and}
  \bibinfo{author}{\bibfnamefont{H.}~\bibnamefont{Takesue}},
  \emph{\bibinfo{title}{Large-scale ising spin network based on degenerate
  optical parametric oscillators}}, \bibinfo{journal}{Nature Photonics}
  \textbf{\bibinfo{volume}{10}}, \bibinfo{pages}{415–419}
  (\bibinfo{year}{2016}{\natexlab{a}}), ISSN \bibinfo{issn}{1749-4893}.

\bibitem[{\citenamefont{Clements et~al.}(2017)\citenamefont{Clements, Renema,
  Wen, Chrzanowski, Kolthammer, and Walmsley}}]{clements:17}
\bibinfo{author}{\bibfnamefont{W.~R.} \bibnamefont{Clements}},
  \bibinfo{author}{\bibfnamefont{J.~J.} \bibnamefont{Renema}},
  \bibinfo{author}{\bibfnamefont{Y.~H.} \bibnamefont{Wen}},
  \bibinfo{author}{\bibfnamefont{H.~M.} \bibnamefont{Chrzanowski}},
  \bibinfo{author}{\bibfnamefont{W.~S.} \bibnamefont{Kolthammer}},
  \bibnamefont{and} \bibinfo{author}{\bibfnamefont{I.~A.}
  \bibnamefont{Walmsley}}, \emph{\bibinfo{title}{Gaussian optical ising
  machines}}, \bibinfo{journal}{Phys. Rev. A} \textbf{\bibinfo{volume}{96}},
  \bibinfo{pages}{043850} (\bibinfo{year}{2017}).

\bibitem[{\citenamefont{Pierangeli et~al.}(2019)\citenamefont{Pierangeli,
  Marcucci, and Conti}}]{pierangeli:19}
\bibinfo{author}{\bibfnamefont{D.}~\bibnamefont{Pierangeli}},
  \bibinfo{author}{\bibfnamefont{G.}~\bibnamefont{Marcucci}}, \bibnamefont{and}
  \bibinfo{author}{\bibfnamefont{C.}~\bibnamefont{Conti}},
  \emph{\bibinfo{title}{Large-scale photonic ising machine by spatial light
  modulation}}, \bibinfo{journal}{Phys. Rev. Lett.}
  \textbf{\bibinfo{volume}{122}}, \bibinfo{pages}{213902}
  (\bibinfo{year}{2019}).

\bibitem[{\citenamefont{Mahboob et~al.}(2016)\citenamefont{Mahboob, Okamoto,
  and Yamaguchi}}]{mahboob:16}
\bibinfo{author}{\bibfnamefont{I.}~\bibnamefont{Mahboob}},
  \bibinfo{author}{\bibfnamefont{H.}~\bibnamefont{Okamoto}}, \bibnamefont{and}
  \bibinfo{author}{\bibfnamefont{H.}~\bibnamefont{Yamaguchi}},
  \emph{\bibinfo{title}{An electromechanical ising hamiltonian}},
  \bibinfo{journal}{Science Advances} \textbf{\bibinfo{volume}{2}},
  \bibinfo{pages}{e1600236} (\bibinfo{year}{2016}).

\bibitem[{\citenamefont{Sheldon et~al.}(2019)\citenamefont{Sheldon, Traversa,
  and Di~Ventra}}]{sheldon:19}
\bibinfo{author}{\bibfnamefont{F.}~\bibnamefont{Sheldon}},
  \bibinfo{author}{\bibfnamefont{F.~L.} \bibnamefont{Traversa}},
  \bibnamefont{and}
  \bibinfo{author}{\bibfnamefont{M.}~\bibnamefont{Di~Ventra}},
  \emph{\bibinfo{title}{Taming a nonconvex landscape with dynamical long-range
  order: Memcomputing ising benchmarks}}, \bibinfo{journal}{Physical Review E}
  \textbf{\bibinfo{volume}{100}} (\bibinfo{year}{2019}), ISSN
  \bibinfo{issn}{2470-0053}.

\bibitem[{\citenamefont{Wang and Roychowdhury}(2019)}]{wang:19a}
\bibinfo{author}{\bibfnamefont{T.}~\bibnamefont{Wang}} \bibnamefont{and}
  \bibinfo{author}{\bibfnamefont{J.}~\bibnamefont{Roychowdhury}},
  \emph{\bibinfo{title}{Oim: Oscillator-based ising machines for solving
  combinatorial optimisation problems}} (\bibinfo{year}{2019}),
  \eprint{1903.07163}.

\bibitem[{\citenamefont{Chou et~al.}(2019)\citenamefont{Chou, Bramhavar, Ghosh,
  and Herzog}}]{chou:19}
\bibinfo{author}{\bibfnamefont{J.}~\bibnamefont{Chou}},
  \bibinfo{author}{\bibfnamefont{S.}~\bibnamefont{Bramhavar}},
  \bibinfo{author}{\bibfnamefont{S.}~\bibnamefont{Ghosh}}, \bibnamefont{and}
  \bibinfo{author}{\bibfnamefont{W.}~\bibnamefont{Herzog}},
  \emph{\bibinfo{title}{{Analog Coupled Oscillator Based Weighted Ising
  Machine}}}, \bibinfo{journal}{{Scientific Reports}}
  \textbf{\bibinfo{volume}{9}}, \bibinfo{pages}{14786} (\bibinfo{year}{2019}),
  ISSN \bibinfo{issn}{2045-2322}.

\bibitem[{\citenamefont{Cai et~al.}(2020)\citenamefont{Cai, Kumar, Vaerenbergh,
  Sheng, Liu, Li, Liu, Foltin, Yu, Xia et~al.}}]{cai:20}
\bibinfo{author}{\bibfnamefont{F.}~\bibnamefont{Cai}},
  \bibinfo{author}{\bibfnamefont{S.}~\bibnamefont{Kumar}},
  \bibinfo{author}{\bibfnamefont{T.~V.} \bibnamefont{Vaerenbergh}},
  \bibinfo{author}{\bibfnamefont{X.}~\bibnamefont{Sheng}},
  \bibinfo{author}{\bibfnamefont{R.}~\bibnamefont{Liu}},
  \bibinfo{author}{\bibfnamefont{C.}~\bibnamefont{Li}},
  \bibinfo{author}{\bibfnamefont{Z.}~\bibnamefont{Liu}},
  \bibinfo{author}{\bibfnamefont{M.}~\bibnamefont{Foltin}},
  \bibinfo{author}{\bibfnamefont{S.}~\bibnamefont{Yu}},
  \bibinfo{author}{\bibfnamefont{Q.}~\bibnamefont{Xia}}, \bibnamefont{et~al.},
  \emph{\bibinfo{title}{Power-efficient combinatorial optimization using
  intrinsic noise in memristor hopfield neural networks}},
  \bibinfo{journal}{Nature Electronics} \textbf{\bibinfo{volume}{3}},
  \bibinfo{pages}{409} (\bibinfo{year}{2020}).

\bibitem[{\citenamefont{Afoakwa et~al.}(2020)\citenamefont{Afoakwa, Zhang,
  Vengalam, Ignjatovic, and Huang}}]{afoakwa:20}
\bibinfo{author}{\bibfnamefont{R.}~\bibnamefont{Afoakwa}},
  \bibinfo{author}{\bibfnamefont{Y.}~\bibnamefont{Zhang}},
  \bibinfo{author}{\bibfnamefont{U.~K.~R.} \bibnamefont{Vengalam}},
  \bibinfo{author}{\bibfnamefont{Z.}~\bibnamefont{Ignjatovic}},
  \bibnamefont{and} \bibinfo{author}{\bibfnamefont{M.}~\bibnamefont{Huang}},
  \emph{\bibinfo{title}{Cmos ising machines with coupled bistable nodes}}
  (\bibinfo{year}{2020}), \eprint{2007.06665}.

\bibitem[{\citenamefont{Berloff et~al.}(2017)\citenamefont{Berloff, Silva,
  Kalinin, Askitopoulos, Töpfer, Cilibrizzi, Langbein, and
  Lagoudakis}}]{berloff:17}
\bibinfo{author}{\bibfnamefont{N.~G.} \bibnamefont{Berloff}},
  \bibinfo{author}{\bibfnamefont{M.}~\bibnamefont{Silva}},
  \bibinfo{author}{\bibfnamefont{K.}~\bibnamefont{Kalinin}},
  \bibinfo{author}{\bibfnamefont{A.}~\bibnamefont{Askitopoulos}},
  \bibinfo{author}{\bibfnamefont{J.~D.} \bibnamefont{Töpfer}},
  \bibinfo{author}{\bibfnamefont{P.}~\bibnamefont{Cilibrizzi}},
  \bibinfo{author}{\bibfnamefont{W.}~\bibnamefont{Langbein}}, \bibnamefont{and}
  \bibinfo{author}{\bibfnamefont{P.~G.} \bibnamefont{Lagoudakis}},
  \emph{\bibinfo{title}{Realizing the classical xy hamiltonian in
  polariton simulators}}, \bibinfo{journal}{Nature Materials}
  \textbf{\bibinfo{volume}{16}}, \bibinfo{pages}{1120–1126}
  (\bibinfo{year}{2017}), ISSN \bibinfo{issn}{1476-4660}.

\bibitem[{\citenamefont{Kalinin and Berloff}(2018{\natexlab{a}})}]{kalinin:18a}
\bibinfo{author}{\bibfnamefont{K.~P.} \bibnamefont{Kalinin}} \bibnamefont{and}
  \bibinfo{author}{\bibfnamefont{N.~G.} \bibnamefont{Berloff}},
  \emph{\bibinfo{title}{Global optimization of spin hamiltonians with
  gain-dissipative systems}} (\bibinfo{year}{2018}{\natexlab{a}}),
  \eprint{1807.00699}.

\bibitem[{\citenamefont{Strogatz}(2015)}]{Strogatz:15}
\bibinfo{author}{\bibfnamefont{S.}~\bibnamefont{Strogatz}},
  \emph{\bibinfo{title}{Nonlinear Dynamics and Chaos: With Applications to
  Physics, Biology, Chemistry, and Engineering}}, Studies in Nonlinearity
  (\bibinfo{publisher}{Westview Press}, \bibinfo{year}{2015}),
  \bibinfo{edition}{2nd} ed., ISBN \bibinfo{isbn}{9780813349107}.

\bibitem[{\citenamefont{Hopfield}(1982)}]{hopfield:82}
\bibinfo{author}{\bibfnamefont{J.~J.} \bibnamefont{Hopfield}},
  \emph{\bibinfo{title}{{{Neural networks and physical systems with emergent
  collective computational abilities}}}}, \bibinfo{journal}{Proc. Natl. Acad.
  Sci.} \textbf{\bibinfo{volume}{79}}, \bibinfo{pages}{2554}
  (\bibinfo{year}{1982}).

\bibitem[{\citenamefont{Hopfield and Tank}(1985)}]{hopfield:85}
\bibinfo{author}{\bibfnamefont{J.~J.} \bibnamefont{Hopfield}} \bibnamefont{and}
  \bibinfo{author}{\bibfnamefont{D.~W.} \bibnamefont{Tank}},
  \emph{\bibinfo{title}{"neural" computation of decisions in optimization
  problems}}, \bibinfo{journal}{Biol. Cybern.} \textbf{\bibinfo{volume}{52}},
  \bibinfo{pages}{141–152} (\bibinfo{year}{1985}), ISSN
  \bibinfo{issn}{0340-1200}.

\bibitem[{\citenamefont{Waugh et~al.}(1990)\citenamefont{Waugh, Marcus, and
  Westervelt}}]{waugh:90}
\bibinfo{author}{\bibfnamefont{F.~R.} \bibnamefont{Waugh}},
  \bibinfo{author}{\bibfnamefont{C.~M.} \bibnamefont{Marcus}},
  \bibnamefont{and} \bibinfo{author}{\bibfnamefont{R.~M.}
  \bibnamefont{Westervelt}}, \emph{\bibinfo{title}{Fixed-point attractors in
  analog neural computation}}, \bibinfo{journal}{Phys. Rev. Lett.}
  \textbf{\bibinfo{volume}{64}}, \bibinfo{pages}{1986} (\bibinfo{year}{1990}).

\bibitem[{\citenamefont{Fukai and Shiino}(1990)}]{fukai:90}
\bibinfo{author}{\bibfnamefont{T.}~\bibnamefont{Fukai}} \bibnamefont{and}
  \bibinfo{author}{\bibfnamefont{M.}~\bibnamefont{Shiino}},
  \emph{\bibinfo{title}{Large suppression of spurious states in neural networks
  of nonlinear analog neurons}}, \bibinfo{journal}{Phys. Rev. A}
  \textbf{\bibinfo{volume}{42}}, \bibinfo{pages}{7459} (\bibinfo{year}{1990}).

\bibitem[{\citenamefont{Wilson and Pawley}(1988)}]{wilson:88}
\bibinfo{author}{\bibfnamefont{G.~V.} \bibnamefont{Wilson}} \bibnamefont{and}
  \bibinfo{author}{\bibfnamefont{G.~S.} \bibnamefont{Pawley}},
  \emph{\bibinfo{title}{On the stability of the travelling salesman problem
  algorithm of hopfield and tank}}, \bibinfo{journal}{Biol. Cybern.}
  \textbf{\bibinfo{volume}{58}}, \bibinfo{pages}{63–70}
  (\bibinfo{year}{1988}), ISSN \bibinfo{issn}{0340-1200}.

\bibitem[{\citenamefont{Leleu et~al.}(2017)\citenamefont{Leleu, Yamamoto,
  Utsunomiya, and Aihara}}]{leleu:17}
\bibinfo{author}{\bibfnamefont{T.}~\bibnamefont{Leleu}},
  \bibinfo{author}{\bibfnamefont{Y.}~\bibnamefont{Yamamoto}},
  \bibinfo{author}{\bibfnamefont{S.}~\bibnamefont{Utsunomiya}},
  \bibnamefont{and} \bibinfo{author}{\bibfnamefont{K.}~\bibnamefont{Aihara}},
  \emph{\bibinfo{title}{Combinatorial optimization using dynamical phase
  transitions in driven-dissipative systems}}, \bibinfo{journal}{Phys. Rev. E}
  \textbf{\bibinfo{volume}{95}}, \bibinfo{pages}{022118}
  (\bibinfo{year}{2017}).

\bibitem[{\citenamefont{{Isakov} et~al.}(2015)\citenamefont{{Isakov},
  {Zintchenko}, {R{\o}nnow}, and {Troyer}}}]{isakov:15}
\bibinfo{author}{\bibfnamefont{S.~V.} \bibnamefont{{Isakov}}},
  \bibinfo{author}{\bibfnamefont{I.~N.} \bibnamefont{{Zintchenko}}},
  \bibinfo{author}{\bibfnamefont{T.~F.} \bibnamefont{{R{\o}nnow}}},
  \bibnamefont{and} \bibinfo{author}{\bibfnamefont{M.}~\bibnamefont{{Troyer}}},
  \emph{\bibinfo{title}{{{Optimized simulated annealing for Ising spin
  glasses}}}}, \bibinfo{journal}{Comput. Phys. Commun.}
  \textbf{\bibinfo{volume}{192}}, \bibinfo{pages}{265} (\bibinfo{year}{2015}),
  \bibinfo{note}{(see also ancillary material to arxiv:cond-mat/1401.1084)}.

\bibitem[{\citenamefont{Sherrington and Kirkpatrick}(1975)}]{sherrington:75}
\bibinfo{author}{\bibfnamefont{D.}~\bibnamefont{Sherrington}} \bibnamefont{and}
  \bibinfo{author}{\bibfnamefont{S.}~\bibnamefont{Kirkpatrick}},
  \emph{\bibinfo{title}{Solvable model of a spin glass}},
  \bibinfo{journal}{Phys. Rev. Lett.} \textbf{\bibinfo{volume}{35}},
  \bibinfo{pages}{1792} (\bibinfo{year}{1975}).

\bibitem[{\citenamefont{Viana and Bray}(1985)}]{viana:85}
\bibinfo{author}{\bibfnamefont{L.}~\bibnamefont{Viana}} \bibnamefont{and}
  \bibinfo{author}{\bibfnamefont{A.~J.} \bibnamefont{Bray}},
  \emph{\bibinfo{title}{Phase diagrams for dilute spin glasses}},
  \bibinfo{journal}{J. Phys. C} \textbf{\bibinfo{volume}{18}},
  \bibinfo{pages}{3037} (\bibinfo{year}{1985}).

\bibitem[{\citenamefont{Goto et~al.}(2021)\citenamefont{Goto, Endo, Suzuki,
  Sakai, Kanao, Hamakawa, Hidaka, Yamasaki, and Tatsumura}}]{goto:21}
\bibinfo{author}{\bibfnamefont{H.}~\bibnamefont{Goto}},
  \bibinfo{author}{\bibfnamefont{K.}~\bibnamefont{Endo}},
  \bibinfo{author}{\bibfnamefont{M.}~\bibnamefont{Suzuki}},
  \bibinfo{author}{\bibfnamefont{Y.}~\bibnamefont{Sakai}},
  \bibinfo{author}{\bibfnamefont{T.}~\bibnamefont{Kanao}},
  \bibinfo{author}{\bibfnamefont{Y.}~\bibnamefont{Hamakawa}},
  \bibinfo{author}{\bibfnamefont{R.}~\bibnamefont{Hidaka}},
  \bibinfo{author}{\bibfnamefont{M.}~\bibnamefont{Yamasaki}}, \bibnamefont{and}
  \bibinfo{author}{\bibfnamefont{K.}~\bibnamefont{Tatsumura}},
  \emph{\bibinfo{title}{High-performance combinatorial optimization based on
  classical mechanics}}, \bibinfo{journal}{Science Advances}
  \textbf{\bibinfo{volume}{7}}, \bibinfo{pages}{eabe7953}
  (\bibinfo{year}{2021}).

\bibitem[{\citenamefont{Inagaki
  et~al.}(2016{\natexlab{b}})\citenamefont{Inagaki, Haribara, Igarashi, Sonobe,
  Tamate, Honjo, Marandi, McMahon, Umeki, Enbutsu et~al.}}]{inagaki:16b}
\bibinfo{author}{\bibfnamefont{T.}~\bibnamefont{Inagaki}},
  \bibinfo{author}{\bibfnamefont{Y.}~\bibnamefont{Haribara}},
  \bibinfo{author}{\bibfnamefont{K.}~\bibnamefont{Igarashi}},
  \bibinfo{author}{\bibfnamefont{T.}~\bibnamefont{Sonobe}},
  \bibinfo{author}{\bibfnamefont{S.}~\bibnamefont{Tamate}},
  \bibinfo{author}{\bibfnamefont{T.}~\bibnamefont{Honjo}},
  \bibinfo{author}{\bibfnamefont{A.}~\bibnamefont{Marandi}},
  \bibinfo{author}{\bibfnamefont{P.~L.} \bibnamefont{McMahon}},
  \bibinfo{author}{\bibfnamefont{T.}~\bibnamefont{Umeki}},
  \bibinfo{author}{\bibfnamefont{K.}~\bibnamefont{Enbutsu}},
  \bibnamefont{et~al.}, \emph{\bibinfo{title}{A coherent ising machine for
  2000-node optimization problems}}, \bibinfo{journal}{Science}
  \textbf{\bibinfo{volume}{354}}, \bibinfo{pages}{603}
  (\bibinfo{year}{2016}{\natexlab{b}}).

\bibitem[{\citenamefont{Yamamoto et~al.}(2020)\citenamefont{Yamamoto, Leleu,
  Ganguli, and Mabuchi}}]{yamamoto:17}
\bibinfo{author}{\bibfnamefont{Y.}~\bibnamefont{Yamamoto}},
  \bibinfo{author}{\bibfnamefont{T.}~\bibnamefont{Leleu}},
  \bibinfo{author}{\bibfnamefont{S.}~\bibnamefont{Ganguli}}, \bibnamefont{and}
  \bibinfo{author}{\bibfnamefont{H.}~\bibnamefont{Mabuchi}},
  \emph{\bibinfo{title}{Coherent ising machines—quantum optics and neural
  network perspectives}}, \bibinfo{journal}{Applied Physics Letters}
  \textbf{\bibinfo{volume}{117}}, \bibinfo{pages}{160501}
  (\bibinfo{year}{2020}).

\bibitem[{\citenamefont{Hamerly et~al.}(2019)\citenamefont{Hamerly, Inagaki,
  McMahon, Venturelli, Marandi, Onodera, Ng, Langrock, Inaba, Honjo
  et~al.}}]{hamerly:19}
\bibinfo{author}{\bibfnamefont{R.}~\bibnamefont{Hamerly}},
  \bibinfo{author}{\bibfnamefont{T.}~\bibnamefont{Inagaki}},
  \bibinfo{author}{\bibfnamefont{P.~L.} \bibnamefont{McMahon}},
  \bibinfo{author}{\bibfnamefont{D.}~\bibnamefont{Venturelli}},
  \bibinfo{author}{\bibfnamefont{A.}~\bibnamefont{Marandi}},
  \bibinfo{author}{\bibfnamefont{T.}~\bibnamefont{Onodera}},
  \bibinfo{author}{\bibfnamefont{E.}~\bibnamefont{Ng}},
  \bibinfo{author}{\bibfnamefont{C.}~\bibnamefont{Langrock}},
  \bibinfo{author}{\bibfnamefont{K.}~\bibnamefont{Inaba}},
  \bibinfo{author}{\bibfnamefont{T.}~\bibnamefont{Honjo}},
  \bibnamefont{et~al.}, \emph{\bibinfo{title}{Experimental investigation of
  performance differences between coherent ising machines and a quantum
  annealer}}, \bibinfo{journal}{Science Advances} \textbf{\bibinfo{volume}{5}},
  \bibinfo{pages}{eaau0823} (\bibinfo{year}{2019}).

\bibitem[{\citenamefont{Haribara et~al.}(2017)\citenamefont{Haribara, Ishikawa,
  Utsunomiya, Aihara, and Yamamoto}}]{haribara:17}
\bibinfo{author}{\bibfnamefont{Y.}~\bibnamefont{Haribara}},
  \bibinfo{author}{\bibfnamefont{H.}~\bibnamefont{Ishikawa}},
  \bibinfo{author}{\bibfnamefont{S.}~\bibnamefont{Utsunomiya}},
  \bibinfo{author}{\bibfnamefont{K.}~\bibnamefont{Aihara}}, \bibnamefont{and}
  \bibinfo{author}{\bibfnamefont{Y.}~\bibnamefont{Yamamoto}},
  \emph{\bibinfo{title}{Performance evaluation of coherent ising machines
  against classical neural networks}}, \bibinfo{journal}{Quantum Science and
  Technology} \textbf{\bibinfo{volume}{2}}, \bibinfo{pages}{044002}
  (\bibinfo{year}{2017}), ISSN \bibinfo{issn}{2058-9565}.

\bibitem[{\citenamefont{Tiunov et~al.}(2019)\citenamefont{Tiunov, Ulanov, and
  Lvovsky}}]{tiunov:19}
\bibinfo{author}{\bibfnamefont{E.~S.} \bibnamefont{Tiunov}},
  \bibinfo{author}{\bibfnamefont{A.~E.} \bibnamefont{Ulanov}},
  \bibnamefont{and} \bibinfo{author}{\bibfnamefont{A.~I.}
  \bibnamefont{Lvovsky}}, \emph{\bibinfo{title}{Annealing by simulating the
  coherent ising machine}}, \bibinfo{journal}{Optics Express}
  \textbf{\bibinfo{volume}{27}}, \bibinfo{pages}{10288} (\bibinfo{year}{2019}),
  ISSN \bibinfo{issn}{1094-4087}.

\bibitem[{\citenamefont{B{\"o}hm et~al.}(2019)\citenamefont{B{\"o}hm,
  Verschaffelt, and Van~der Sand}}]{boehm:19}
\bibinfo{author}{\bibfnamefont{F.}~\bibnamefont{B{\"o}hm}},
  \bibinfo{author}{\bibfnamefont{G.}~\bibnamefont{Verschaffelt}},
  \bibnamefont{and} \bibinfo{author}{\bibfnamefont{G.}~\bibnamefont{Van~der
  Sand}}, \emph{\bibinfo{title}{A poor man’s coherent ising machine based on
  opto-electronic feedback systems for solving optimization problems}},
  \bibinfo{journal}{Nature Communications} \textbf{\bibinfo{volume}{10}},
  \bibinfo{pages}{3538} (\bibinfo{year}{2019}), ISSN \bibinfo{issn}{2041-1723}.

\bibitem[{\citenamefont{Kalinin and Berloff}(2018{\natexlab{b}})}]{kalinin:18b}
\bibinfo{author}{\bibfnamefont{K.~P.} \bibnamefont{Kalinin}} \bibnamefont{and}
  \bibinfo{author}{\bibfnamefont{N.~G.} \bibnamefont{Berloff}},
  \emph{\bibinfo{title}{Networks of non-equilibrium condensates for global
  optimization}}, \bibinfo{journal}{New Journal of Physics}
  \textbf{\bibinfo{volume}{20}}, \bibinfo{pages}{113023}
  (\bibinfo{year}{2018}{\natexlab{b}}), ISSN \bibinfo{issn}{1367-2630}.

\bibitem[{\citenamefont{Landau and Lifshitz}(1980)}]{landau:80}
\bibinfo{author}{\bibfnamefont{L.~D.} \bibnamefont{Landau}} \bibnamefont{and}
  \bibinfo{author}{\bibfnamefont{E.~M.} \bibnamefont{Lifshitz}},
  \emph{\bibinfo{title}{{Statistical Physics, Part 1}}},
  vol.~\bibinfo{volume}{5} of \emph{\bibinfo{series}{Course of Theoretical
  Physics}} (\bibinfo{publisher}{Butterworth-Heinemann}, \bibinfo{year}{1980}),
  \bibinfo{edition}{3rd} ed.

\bibitem[{\citenamefont{Yeomans}(1992)}]{yeomans:92}
\bibinfo{author}{\bibfnamefont{J.~M.} \bibnamefont{Yeomans}},
  \emph{\bibinfo{title}{{Statistical Mechanics of Phase Transitions}}}
  (\bibinfo{publisher}{Oxford University Press}, \bibinfo{address}{Oxford},
  \bibinfo{year}{1992}).

\bibitem[{\citenamefont{Stroev and Berloff}(2021)}]{Kalinin:21}
\bibinfo{author}{\bibfnamefont{N.}~\bibnamefont{Stroev}} \bibnamefont{and}
  \bibinfo{author}{\bibfnamefont{N.~G.} \bibnamefont{Berloff}},
  \emph{\bibinfo{title}{Discrete polynomial optimization with coherent networks
  of condensates and complex coupling switching}}, \bibinfo{journal}{Physical
  Review Letters} \textbf{\bibinfo{volume}{126}} (\bibinfo{year}{2021}), ISSN
  \bibinfo{issn}{1079-7114}.

\bibitem[{\citenamefont{Harrison et~al.}(2020)\citenamefont{Harrison,
  Sigurdsson, and Lagoudakis}}]{harrison:20}
\bibinfo{author}{\bibfnamefont{S.~L.} \bibnamefont{Harrison}},
  \bibinfo{author}{\bibfnamefont{H.}~\bibnamefont{Sigurdsson}},
  \bibnamefont{and} \bibinfo{author}{\bibfnamefont{P.~G.}
  \bibnamefont{Lagoudakis}}, \emph{\bibinfo{title}{Solving the max-3-cut
  problem using synchronized dissipative networks}} (\bibinfo{year}{2020}),
  \eprint{2007.06135}.

\bibitem[{\citenamefont{Wang et~al.}(2019)\citenamefont{Wang, Wu, and
  Roychowdhury}}]{wang:19b}
\bibinfo{author}{\bibfnamefont{T.}~\bibnamefont{Wang}},
  \bibinfo{author}{\bibfnamefont{L.}~\bibnamefont{Wu}}, \bibnamefont{and}
  \bibinfo{author}{\bibfnamefont{J.}~\bibnamefont{Roychowdhury}},
  \emph{\bibinfo{title}{Late breaking results: New computational results and
  hardware prototypes for oscillator-based ising machines}}
  (\bibinfo{year}{2019}), \eprint{1904.10211}.

\bibitem[{\citenamefont{Goto}(2016)}]{goto:16}
\bibinfo{author}{\bibfnamefont{H.}~\bibnamefont{Goto}},
  \emph{\bibinfo{title}{Bifurcation-based adiabatic quantum computation with a
  nonlinear oscillator network}}, \bibinfo{journal}{Scientific Reports}
  \textbf{\bibinfo{volume}{6}} (\bibinfo{year}{2016}), ISSN
  \bibinfo{issn}{2045-2322}.

\bibitem[{\citenamefont{Nigg et~al.}(2017)\citenamefont{Nigg, Lörch, and
  Tiwari}}]{nigg:17}
\bibinfo{author}{\bibfnamefont{S.~E.} \bibnamefont{Nigg}},
  \bibinfo{author}{\bibfnamefont{N.}~\bibnamefont{Lörch}}, \bibnamefont{and}
  \bibinfo{author}{\bibfnamefont{R.~P.} \bibnamefont{Tiwari}},
  \emph{\bibinfo{title}{Robust quantum optimizer with full connectivity}},
  \bibinfo{journal}{Science Advances} \textbf{\bibinfo{volume}{3}}
  (\bibinfo{year}{2017}), ISSN \bibinfo{issn}{2375-2548}.

\bibitem[{\citenamefont{Goto et~al.}(2019)\citenamefont{Goto, Tatsumura, and
  Dixon}}]{goto:19}
\bibinfo{author}{\bibfnamefont{H.}~\bibnamefont{Goto}},
  \bibinfo{author}{\bibfnamefont{K.}~\bibnamefont{Tatsumura}},
  \bibnamefont{and} \bibinfo{author}{\bibfnamefont{A.~R.} \bibnamefont{Dixon}},
  \emph{\bibinfo{title}{Combinatorial optimization by simulating adiabatic
  bifurcations in nonlinear hamiltonian systems}}, \bibinfo{journal}{Science
  Advances} \textbf{\bibinfo{volume}{5}}, \bibinfo{pages}{eaav2372}
  (\bibinfo{year}{2019}).

\bibitem[{\citenamefont{Du and Swamy}(2014)}]{du:14}
\bibinfo{author}{\bibfnamefont{K.-L.} \bibnamefont{Du}} \bibnamefont{and}
  \bibinfo{author}{\bibfnamefont{M.~N.~S.} \bibnamefont{Swamy}},
  \emph{\bibinfo{title}{{Neural Networks and Statistical Learning}}}
  (\bibinfo{publisher}{Springer}, \bibinfo{address}{London},
  \bibinfo{year}{2014}), ISBN \bibinfo{isbn}{9781447155713}.

\bibitem[{\citenamefont{Hopfield}(1984)}]{hopfield:84}
\bibinfo{author}{\bibfnamefont{J.~J.} \bibnamefont{Hopfield}},
  \emph{\bibinfo{title}{Neurons with graded response have collective
  computational properties like those of two-state neurons}},
  \bibinfo{journal}{Proceedings of the National Academy of Sciences}
  \textbf{\bibinfo{volume}{81}}, \bibinfo{pages}{3088} (\bibinfo{year}{1984}),
  ISSN \bibinfo{issn}{0027-8424}.

\bibitem[{\citenamefont{Hopfield and Tank}(1986)}]{hopfield:86}
\bibinfo{author}{\bibfnamefont{J.~J.} \bibnamefont{Hopfield}} \bibnamefont{and}
  \bibinfo{author}{\bibfnamefont{D.~W.} \bibnamefont{Tank}},
  \emph{\bibinfo{title}{Computing with neural circuits: A model}},
  \bibinfo{journal}{Science} \textbf{\bibinfo{volume}{233}},
  \bibinfo{pages}{625} (\bibinfo{year}{1986}).

\bibitem[{\citenamefont{Qian}(1999)}]{qian:99}
\bibinfo{author}{\bibfnamefont{N.}~\bibnamefont{Qian}},
  \emph{\bibinfo{title}{On the momentum term in gradient descent learning
  algorithms}}, \bibinfo{journal}{Neural Networks}
  \textbf{\bibinfo{volume}{12}}, \bibinfo{pages}{145} (\bibinfo{year}{1999}),
  ISSN \bibinfo{issn}{0893-6080}.

\bibitem[{\citenamefont{Perera et~al.}(2021)\citenamefont{Perera, Akpabio,
  Hamze, Mandra, Rose, Aramon, and Katzgraber}}]{perera:21}
\bibinfo{author}{\bibfnamefont{D.}~\bibnamefont{Perera}},
  \bibinfo{author}{\bibfnamefont{I.}~\bibnamefont{Akpabio}},
  \bibinfo{author}{\bibfnamefont{F.}~\bibnamefont{Hamze}},
  \bibinfo{author}{\bibfnamefont{S.}~\bibnamefont{Mandra}},
  \bibinfo{author}{\bibfnamefont{N.}~\bibnamefont{Rose}},
  \bibinfo{author}{\bibfnamefont{M.}~\bibnamefont{Aramon}}, \bibnamefont{and}
  \bibinfo{author}{\bibfnamefont{H.~G.} \bibnamefont{Katzgraber}},
  \emph{\bibinfo{title}{Chook -- a comprehensive suite for generating binary
  optimization problems with planted solutions}} (\bibinfo{year}{2021}),
  \eprint{2005.14344}.

\bibitem[{\citenamefont{Hamze et~al.}(2018)\citenamefont{Hamze, Jacob, Ochoa,
  Perera, Wang, and Katzgraber}}]{hamze:18}
\bibinfo{author}{\bibfnamefont{F.}~\bibnamefont{Hamze}},
  \bibinfo{author}{\bibfnamefont{D.~C.} \bibnamefont{Jacob}},
  \bibinfo{author}{\bibfnamefont{A.~J.} \bibnamefont{Ochoa}},
  \bibinfo{author}{\bibfnamefont{D.}~\bibnamefont{Perera}},
  \bibinfo{author}{\bibfnamefont{W.}~\bibnamefont{Wang}}, \bibnamefont{and}
  \bibinfo{author}{\bibfnamefont{H.~G.} \bibnamefont{Katzgraber}},
  \emph{\bibinfo{title}{{From near to eternity: Spin-glass planting, tiling
  puzzles, and constraint-satisfaction problems}}}, \bibinfo{journal}{Phys.
  Rev. E} \textbf{\bibinfo{volume}{97}}, \bibinfo{pages}{043303}
  (\bibinfo{year}{2018}).

\bibitem[{\citenamefont{Perera et~al.}(2020)\citenamefont{Perera, Hamze,
  Raymond, Weigel, and Katzgraber}}]{perera:20}
\bibinfo{author}{\bibfnamefont{D.}~\bibnamefont{Perera}},
  \bibinfo{author}{\bibfnamefont{F.}~\bibnamefont{Hamze}},
  \bibinfo{author}{\bibfnamefont{J.}~\bibnamefont{Raymond}},
  \bibinfo{author}{\bibfnamefont{M.}~\bibnamefont{Weigel}}, \bibnamefont{and}
  \bibinfo{author}{\bibfnamefont{H.~G.} \bibnamefont{Katzgraber}},
  \emph{\bibinfo{title}{Computational hardness of spin-glass problems with
  tile-planted solutions}}, \bibinfo{journal}{Phys. Rev. E}
  \textbf{\bibinfo{volume}{101}}, \bibinfo{pages}{023316}
  (\bibinfo{year}{2020}).

\bibitem[{\citenamefont{{R{\o}nnow} et~al.}(2014)\citenamefont{{R{\o}nnow},
  {Wang}, {Job}, {Boixo}, {Isakov}, {Wecker}, {Martinis}, {Lidar}, and
  {Troyer}}}]{ronnow:14a}
\bibinfo{author}{\bibfnamefont{T.~F.} \bibnamefont{{R{\o}nnow}}},
  \bibinfo{author}{\bibfnamefont{Z.}~\bibnamefont{{Wang}}},
  \bibinfo{author}{\bibfnamefont{J.}~\bibnamefont{{Job}}},
  \bibinfo{author}{\bibfnamefont{S.}~\bibnamefont{{Boixo}}},
  \bibinfo{author}{\bibfnamefont{S.~V.} \bibnamefont{{Isakov}}},
  \bibinfo{author}{\bibfnamefont{D.}~\bibnamefont{{Wecker}}},
  \bibinfo{author}{\bibfnamefont{J.~M.} \bibnamefont{{Martinis}}},
  \bibinfo{author}{\bibfnamefont{D.~A.} \bibnamefont{{Lidar}}},
  \bibnamefont{and} \bibinfo{author}{\bibfnamefont{M.}~\bibnamefont{{Troyer}}},
  \emph{\bibinfo{title}{{Defining and detecting quantum speedup}}},
  \bibinfo{journal}{Science} \textbf{\bibinfo{volume}{345}},
  \bibinfo{pages}{420} (\bibinfo{year}{2014}).

\bibitem[{\citenamefont{Caparr{\'o}s et~al.}(2002)\citenamefont{Caparr{\'o}s,
  Ruiz, and Hern{\'a}ndez}}]{joya:02}
\bibinfo{author}{\bibfnamefont{G.~J.} \bibnamefont{Caparr{\'o}s}},
  \bibinfo{author}{\bibfnamefont{M.~A.~A.} \bibnamefont{Ruiz}},
  \bibnamefont{and} \bibinfo{author}{\bibfnamefont{F.~S.}
  \bibnamefont{Hern{\'a}ndez}}, \emph{\bibinfo{title}{Hopfield neural networks
  for optimization: study of the different dynamics}},
  \bibinfo{journal}{Neurocomputing} \textbf{\bibinfo{volume}{43}},
  \bibinfo{pages}{219} (\bibinfo{year}{2002}).

\bibitem[{\citenamefont{Liu and Wang}(2009)}]{liu:09}
\bibinfo{author}{\bibfnamefont{W.}~\bibnamefont{Liu}} \bibnamefont{and}
  \bibinfo{author}{\bibfnamefont{L.}~\bibnamefont{Wang}},
  \emph{\bibinfo{title}{Solving the shortest path routing problem using noisy
  hopfield neural networks}}, \bibinfo{journal}{2009 WRI International
  Conference on Communications and Mobile Computing}
  \textbf{\bibinfo{volume}{2}}, \bibinfo{pages}{299} (\bibinfo{year}{2009}).

\bibitem[{\citenamefont{Zhang and Di~Ventra}(2021)}]{zhang:21}
\bibinfo{author}{\bibfnamefont{Y.-H.} \bibnamefont{Zhang}} \bibnamefont{and}
  \bibinfo{author}{\bibfnamefont{M.}~\bibnamefont{Di~Ventra}},
  \emph{\bibinfo{title}{Directed percolation and numerical stability of
  simulations of digital memcomputing machines}}, \bibinfo{journal}{Chaos: An
  Interdisciplinary Journal of Nonlinear Science}
  \textbf{\bibinfo{volume}{31}}, \bibinfo{pages}{063127}
  (\bibinfo{year}{2021}), ISSN \bibinfo{issn}{1089-7682}.

\bibitem[{\citenamefont{Andresen et~al.}(2013)\citenamefont{Andresen, Zhu,
  Andrist, Katzgraber, Dobrosavljevi{\'{c}}, and Zimanyi}}]{andresen:13}
\bibinfo{author}{\bibfnamefont{J.~C.} \bibnamefont{Andresen}},
  \bibinfo{author}{\bibfnamefont{Z.}~\bibnamefont{Zhu}},
  \bibinfo{author}{\bibfnamefont{R.~S.} \bibnamefont{Andrist}},
  \bibinfo{author}{\bibfnamefont{H.~G.} \bibnamefont{Katzgraber}},
  \bibinfo{author}{\bibfnamefont{V.}~\bibnamefont{Dobrosavljevi{\'{c}}}},
  \bibnamefont{and} \bibinfo{author}{\bibfnamefont{G.~T.}
  \bibnamefont{Zimanyi}}, \emph{\bibinfo{title}{{Self-Organized Criticality in
  Glassy Spin Systems Requires a Diverging Number of Neighbors}}},
  \bibinfo{journal}{Phys. Rev. Lett.} \textbf{\bibinfo{volume}{111}},
  \bibinfo{pages}{097203} (\bibinfo{year}{2013}).

\bibitem[{\citenamefont{{Lanting} et~al.}(2014)\citenamefont{{Lanting},
  {Przybysz}, {Smirnov}, {Spedalieri}, {Amin}, {Berkley}, {Harris}, {Altomare},
  {Boixo}, {Bunyk} et~al.}}]{lanting:14}
\bibinfo{author}{\bibfnamefont{T.}~\bibnamefont{{Lanting}}},
  \bibinfo{author}{\bibfnamefont{A.~J.} \bibnamefont{{Przybysz}}},
  \bibinfo{author}{\bibfnamefont{A.~Y.} \bibnamefont{{Smirnov}}},
  \bibinfo{author}{\bibfnamefont{F.~M.} \bibnamefont{{Spedalieri}}},
  \bibinfo{author}{\bibfnamefont{M.~H.} \bibnamefont{{Amin}}},
  \bibinfo{author}{\bibfnamefont{A.~J.} \bibnamefont{{Berkley}}},
  \bibinfo{author}{\bibfnamefont{R.}~\bibnamefont{{Harris}}},
  \bibinfo{author}{\bibfnamefont{F.}~\bibnamefont{{Altomare}}},
  \bibinfo{author}{\bibfnamefont{S.}~\bibnamefont{{Boixo}}},
  \bibinfo{author}{\bibfnamefont{P.}~\bibnamefont{{Bunyk}}},
  \bibnamefont{et~al.}, \emph{\bibinfo{title}{Entanglement in a quantum
  annealing processor}}, \bibinfo{journal}{Phys. Rev. X}
  \textbf{\bibinfo{volume}{4}}, \bibinfo{pages}{021041} (\bibinfo{year}{2014}).

\end{thebibliography}

\end{document}